\documentclass[fp,twocolumn]{jpsj3}
\usepackage{txfonts}
\usepackage{multirow}

\title{4D-XY Superfluid Transition and Dissipation in $^4$He Confined in Nanoporous Media}

\author{Tomoyuki Tani$^1$, Yusuke Nago$^1$, Satoshi Murakawa$^2$, and Keiya Shirahama$^1$}
\inst{$^1$Department of Physics, Keio University, Yokohama 223-8522, Japan \\
$^2$Cryogenic Research Center, The University of Tokyo, Bunkyo-ku, Tokyo 113-0032, Japan} %\\

\abst{
$^4$He confined in nanoporous Gelsil glass is a unique, strongly correlated Bose system exhibiting quantum phase transition (QPT) by controlling pressure. 
Previous studies revealed that the QPT occurs with four - dimensional (4D) XY criticality, which appears in the zero-temperature limit of the superfluid density. 
However, the $P-T$ phase diagram also suggested that 4D XY nature appears at finite temperatures. 
Here, we have determined the critical exponent of the superfluid density of $^4$He in two Gelsil samples that have pore diameter to be about 3 nm, using a newly developed mechanical resonator technique. 
The critical exponent $\zeta$ in the powerlaw fitting $\rho_{\mathrm s} \propto \left| 1 - T/T_{\mathrm c} \right| ^{\zeta}$, where $T_{\mathrm c}$ is the superfluid transition temperature, was found to be 1.0 $\pm$ 0.1 for all pressures realized in this experiment, 0.1 $<$ $P$ $<$ 2.4 MPa. 
This value of $\zeta$ gives a decisive evidence that the finite-temperature superfluid transition belongs to 4D XY universality class. 
The emergence of the 4D XY criticality is explained by the existence of many nanoscale superfluid droplets, the so called localized Bose - Einstein condensates (LBECs), above $T_{\mathrm c}$. 
Due to the large energy cost for $^4$He atoms to move between the LBECs, the phase of the LBEC order parameters fluctuates not only in spatial (3D) but imaginary time ($+1$D) dimensions, resulting in the 4D XY criticality by a temperature near $T_{\mathrm c}$, which is determined by the finite size of the system in the imaginary time dimension. %, $L_{\tau}$. 
Below $T_{\mathrm c}$, macroscopic superfluidity grows in the nanopores of Gelsil by the alignment of the phases of the LBEC order parameters. 
An excess dissipation peak observed below $T_{\mathrm c}$ is well explained by this phase matching process.
}

\begin{document}
\maketitle

\section{Introduction}\label{Introduction}

In the last decades, quantum phase transitions (QPTs) have been of much interest in condensed matter physics\cite{SondhiRMP1997,SachdevQPT}. 
It has been realized that emergence of QPT is a characteristic of strongly correlated systems\cite{GegenwartNP2008,KeimerN2015}, 
as the strong inter-particle correlation produces quantum fluctuations.
Second-order QPT is characterized by a zero-temperature quantum critical point (QCP), which emerges at a particular value of external parameters. %, such as pressure, applied magnetic field, and system disorder. 
In general, QCP is connected to the classical phase transition line at finite temperatures.  
At 0 K, QPT is driven not only by quantum fluctuations in spatial dimensions but also by those in an imaginary-time dimension $i \tau$, where $\tau$ is $\hbar \beta = \hbar/k_{\mathrm B}T$. 
This suggests that a $d$-dimensional classical phase transition at finite temperatures can be continuously connected to a $(d + z)$-dimensional QPT at 0 K, where $z$ is the dynamical critical exponent\cite{SondhiRMP1997}. 
In the case of $z = 1$, one may expect that a three-dimensional (3D) classical phase transition is terminated by a 4D QPT. 
However, 4D QPT was suggested only in two real systems, underdoped cuprate superconductors\cite{BrounPRL2007,FranzPRL2006} and $^4$He confined in nanoporous media.%\cite{YamamotoPRL2004,YamamotoPRL2008,ShirahamaJLTP2007,ShirahamaLTP2008,ShirahamaJPSJ2008,EggelPRB2011}. 

$^4$He confined in nanoporous media is a strongly correlated bosonic system showing QPT\cite{YamamotoPRL2004,YamamotoPRL2008,ShirahamaJLTP2007,ShirahamaLTP2008,ShirahamaJPSJ2008,EggelPRB2011,EggelThesis2011}. 
At $T_{\lambda} = 2.17$ K and at SVP (near zero pressure), bulk liquid $^4$He undergoes the superfluid $\lambda$ transition, which appears as a critical phenomenon with the 3D XY universality class (O(2) symmetry)\cite{AhlersRMP1980,BarmatzRMP2007}. 
When liquid $^4$He is confined in a porous Gelsil glass with a pore diameter of about 3 nm, the superfluid transition temperature $T_{\mathrm c}$ decreases down to 1.4 K near SVP\cite{YamamotoPRL2004}. 
As the pressure increases,  $T_{\mathrm c}$ and the superfluid density $\rho_{\mathrm s}$ decrease. 
$T_{\mathrm c}$ eventually reaches 0 K at a critical pressure $P_{\mathrm c} \sim 3.3$ MPa, meaning that a QPT occurs at $P_{\mathrm c}$. 
%Such strong suppression of superfluidity was never observed in the previous experiments using porous Vycor glass that has larger nanopores with 6 - 10 nm pore sizes. 
The suppression of superfluidity is attributed to the emergence of a non-trivial nonsuperfluid state consisting of the localized Bose-Einstein condensates (LBECs). 
At a temperature $T_{\mathrm {cp}}$, which is slightly lower than $T_{\lambda}$, heat capacity shows a peak, indicating the superfluid transition in nanopores\cite{YamamotoPRL2008}. 
However, no superfluidity is observed between $T_{\mathrm {cp}}$ and $T_{\mathrm {c}}$. 
The heat capacity peak is attributed to the emergence of many nanoscale LBECs in the pore voids. % below $T_{\mathrm {cp}}$. 
Strong correlation among helium atoms and the confinement within narrow pores suppress the spatial exchange of atoms and hence phase coherence among LBECs, so the system does not show macroscopic superfluidity. 
This trend is reinforced as the pressure, i.e. the correlation in He atoms, increases, eventually resulting in the QPT at $P_{\mathrm c}$.

Eggel et al.\cite{EggelPRB2011,EggelThesis2011} theoretically studied the QPT using the disordered quantum rotor (Bose--Hubbard) model\cite{FisherPRB1989}, in which the LBECs are located at the 3D lattice sites. 
They noticed that the particle-hole (p-h) symmetry in the system is recovered by the randomness in the size of LBEC.
The p-h symmetry keeps $z$ to be unity, thus the QPT belongs to the 4D XY universality class.
Near the QCP, the zero-temperature superfluid density $\rho_{\mathrm s}(T = 0 {\mathrm K})$, which was obtained by extrapolation of $\rho_{\mathrm s}$ measured at finite $T$,  varies as $\rho_{\mathrm s}(T = 0 {\mathrm K}) \propto (P_{\mathrm c} - P)^\zeta$. 
The critical exponent $\zeta$ is given by $\zeta = (d + z - 2)\nu$, where $\nu$ is the critical exponent of the correlation length $\xi$. 
In 4D XY, $\nu = 1/2$, so $\zeta = 1$ and $\rho_{\mathrm s}(T = 0 {\mathrm K}) \propto P_{\mathrm c} - P$. 
This agrees well with the experimental observation that the zero-temperature superfluid density is proportional to the pressure except for the very vicinity of $P_{\mathrm c}$. 
Moreover, the 4D XY criticality shows an unexpected agreement with the experimental result at finite temperatures. 
The theory proposed the temperature dependence of the critical pressure at $T_{\mathrm c}$ as $P_{\mathrm c}(0) - P_{\mathrm c}(T) \propto T^{1/z\nu}$. 
The 4D XY criticality gives $P_{\mathrm c}(0) - P_{\mathrm c}(T) \propto T^2$.
This also agreed well with the experimental result that the phase boundary between the LBEC state and the superfluid state is fitted to a power law $P_{\mathrm c}(0) - P_{\mathrm c}(T) \propto T^{2.13}$.

In the general theory of QPT\cite{SondhiRMP1997}, however, a zero-temperature QCP is altered to a classical critical line at finite temperatures, in this case, a 3D XY superfluid transition. Therefore, the agreements of the experimental observations with the theoretical results of the 4D XY criticality not only at $T = 0$ but at finite temperatures are rather surprising, and motivate us to study the superfluid transition of $^4$He in the nanoporous Gelsil in more detail. 

In the previous work\cite{YamamotoPRL2004}, we employed the torsional oscillator (TO) technique to examine superfluidity of $^4$He in Gelsil.
However, it was difficult to determine the critical exponent of the superfluid density, because the bulk liquid $^4$He that exists in the TO bob (around the Gelsil sample) and in the torsion rod might contribute to the superfluid signal.  
We have developed a superfluid mechanical resonator\cite{AvenelPRL1985,RojasPRB2015} to measure the superflow and dissipation of $^4$He inside the Gelsil.
This method enables us to eliminate the unwanted contribution outside the Gelsil sample and extract only the effect of macroscopic superfluidity in the nanoporous network. 
In the previous letter, we reported on the experimental determination of the critical exponent of the superfluid density near $T_{\mathrm c}$\cite{TaniJPSJ2021}. 
We have found that the the superfluid critical exponent $\zeta$ is $1.0 \pm 0.1$, which agrees with the exponent of the 4D XY universality class.
In this paper, we give a full account on the experiment. 
We have measured dissipation in the superflow characteristics. 
The dissipation below $T_{\mathrm c}$ is explained by the phase matching among the LBECs, which is a necessary process for macroscopic superfluidity throughout the Gelsil sample.  
We also present a scaling analysis for the superfluid density, which was first proposed by Eggel\cite{EggelThesis2011}. 
These analyses further support the finite-temperature 4D XY criticality and the LBEC conjecture.

\section{Experimental}\label{Experimental}
\subsection{Double-diaphragm superfluid mechanical resonator}\label{Resonator}
\begin{figure}[tb]
\centering
\includegraphics[width=0.95\linewidth]{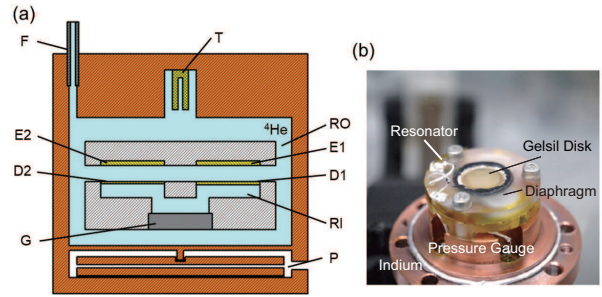}
\caption{\label{HRcell}(Color online)
The double-diaphragm superfluid mechanical resonator. (a) A schematic cross-sectional view. 
The orange parts form an enclosure made of copper.  The gray ones are the resonator body made of Stycast 1266 epoxy.
The abbreviations indicate respectively:
(D1) drive diaphragm, (D2) detector diaphragm, (E1) drive electrode, (E2) detector electrode, (G) porous Gelsil glass disk, (T) quartz tuning fork viscometer, (P) capacitance pressure gauge and (F) liquid filling line. 
Liquid ${}^{\rm 4}$He is filled in inner and outer volumes, and act as helium reservoirs RI and RO. 
(b) Photograph of the resonator. 
The Gelsil glass disk is glued to the macor ring with Stycast 1266 epoxy, in order to match the thermal contraction. 
The macor ring is further glued to the resonator body by Stycast 2850 epoxy, which is shown as black ring. 
One of the diaphragms is seen faintly through the epoxy body.
A copper cap (not shown) is screwed to the copper base shown at the bottom with an indium ring for sealing.
}
\end{figure}

In order to eliminate the effect of crosstalk which may contribute to the background of the signal, we have developed a mechanical resonator that has two flexible diaphragms, in which the drive and detection of the superflow are made separately. 
The experimental apparatus is depicted in Fig.~\ref{HRcell}. 
The superfluid mechanical resonator consists of two liquid $^4$He resorviors RI and RO that are connected by a porous Gelsil glass disk (G).  
%At the bottom of the resonator, a porous gelsil disk (G), which acts as a bottleneck, is glued.
The top wall of the resorvoir RI has two circular diaphragms D1 and D2.
D1 and D2 are metal-deposited Kapton films with different thicknesses, 50 $\rm\mu m$ and 7.5 $\rm\mu m$, respectively. 
Two circular fixed electrodes E1 and E2 are located oppositing to D1 and D2 respectively. 
The filling line F is connected to the gas handling system at room temperature via a metal capillary. 
We have used the first grade (99.999\%) $^4$He gas with natural abundance (perhaps 0.3 ppm $^3$He is contained) for the liquid helium samples. 

%A unique feature of this superfluid resonator is the double diaphragms, which were not employed in the previous superflow experiments using similar resonators \cite{AvenelPRL1985,RojasPRB2015}. 
In the measurement, two volumes RI and RO are filled with liquid $^4$He, so that the bulk liquid in reservoir RI is connected to the liquid in RO via the liquid in Gelsil (G). 
The whole system forms a kind of Helmholtz resonator in which the porous Gelsil acts as a bottleneck, but we have found that the resonances observed in this apparatus cannot be assigned to be the Helmholtz resonance, except for a mode at low frequency (the $R_1$ mode,  see below).
In order to drive and detect the mechanical oscillation, a DC bias voltage of 350 V is applied to both diaphragms. 
An AC voltage from a wave generator is superimposed to the driver electrode E1.
This produces mechanical oscillation of the stiffer diaphragm D1 with the same frequency of AC voltage. 
The oscillation of D1 then mechanically drives liquid $^4$He motion in inner resorvoir RI. 
The oscillation propagates to the detector diaphragm D2 via the liquid in RI.
The motion of D2 induces displacement current on detector electrode E2, which is picked up using a current preamplifier and a lock-in amplifier.

In the resonator cell, a capacitive pressure gauge (P) is installed for precise pressure measurement of bulk liquid $^4$He. 
A quartz tuning fork (T) is also placed in the top of resorvior RO. 
It works as a liquid viscometer, which is helpful in analyzing dissipation of mechanical resonator.
All the measurements were performed in the temperature range between 0.7 and 2.4 K using a cryogen-free $^3$He refrigerator. 
The resonator is connected to the $^3$He stage (pot) via a superconducting heat switch made of lead. 
The temperature of the resonator cell is stabilized by controlling the temperatures of the $^3$He stage and the cell, and by switching on or off the heat switch depending on the target temperature.  

\subsection{Porous Gelsil glass}\label{Gelsil}
\begin{figure}[tb]
\centering
\includegraphics[width=0.95\linewidth]{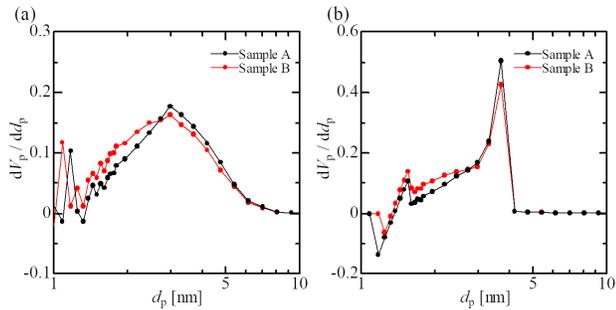}
\caption{(Color online)\label{isotherms} 
The distribution of pore diameters for two samples A and B, derived by applying the BJH method to nitrogen isotherms. 
(a) Distributions obtained from the adsorption isotherms, and (b) from the desorption ones.}
\end{figure}

Gelsil is a nanoporous glass manufactured by the sol-gel method. 
The Gelsil samples we employed in this work were produced by a different manufacturer from that in the previous TO and heat capacity studies\cite{YamamotoPRL2004,YamamotoPRL2008}. 
We have used two Gelsil samples, denoted by sample A and B. 

In Fig.~\ref{isotherms}, we show the pore diameter distribution of the two samples obtained by the BJH method \cite{BarrettJACS1951} applied to nitrogen adsorption and desorption isotherms measured by the Belsorp Mini II apparatus.
Although the nominal pore diameters are 2.5 nm and 3.0 nm for sample A and B, respectively, the pore size distributions measured by us are almost identical. 
The peaks of the distribution are located at 3.0 nm and 3.8 nm, for adsorption and desorption isotherms, respectively. 
The sharp peaks from the desorption isotherms correspond to the size of bottlenecks in the porous structure, while the distribution from adsorption isotherm show the volume of the pores of particular diameters. 
We note that sample B has a slightly larger pore volume in the size range 1.5 $<$ d $<$ 3.0 nm compared to sample A. 
The dimensions of these two disks are 9.0 mm in diameter and 1.0 and 2.0 mm in thickness for sample A and B, respectively. 

\begin{figure}[tb]
\centering
\includegraphics[width=0.95\linewidth]{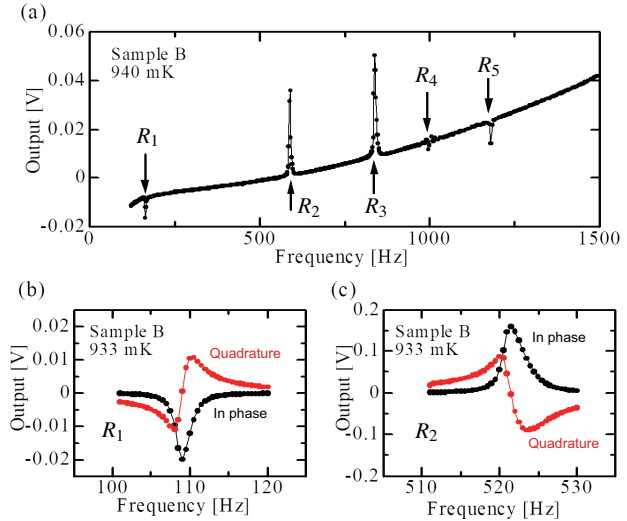}
\caption{\label{HRresonance}
(Color online)
(a) An example of the frequency spectrum of mechanical resonator for sample B. 
Data are the in-phase component, and were taken at 940 mK. 
Five resonances are shown by arrows and denoted as $R_n$ in the frequency order.
(b)(c) In-phase and quadrature components of the (b) $R_1$ and (c) $R_2$ modes taken at 933 mK. 
 }
\end{figure}

\subsection{Experimental procedure}\label{Experiment}
After the $^3$He refrigerator was cooled down, we slowly injected $^4$He to the resonator cell keeping the rate of pressure change in the resorvoir as small as possible.
This is because the resorvoir RI is connected to RO only through the Gelsil disk, which might have a high flow impedance, and the resorvoir wall is made of thin Kapton diaphragms that can easily be broken by large pressure impulse. 
The measurement was limited at the pressure range $0  < P < 2.5$ MPa, above which bulk $^4$He solidifies in the resorvoirs.

In the measurement, we initially measured the frequency spectra of the in-phase and quadrature of the lock-in outputs by sweeping the driving frequency from 10 to 2000 Hz, at some fixed temperatures. 
An example of the spectrum for sample B is shown in Fig.~\ref{HRresonance}(a).
We have observed five sharp resonances at frequencies between 10 and 1250 Hz for both samples A and B. 
As is discussed in the later section, these resonances are determined by the geometry of the reservoir RI, hydrodynamic properties of liquid helium in RI, and superfluid properties of helium in Gelsil that hydrodynamically connects RI and RO.
We denote the resonance modes at frequencies $f_{1, 2, 3, 4, 5}$ in frequency order as $R_{1, 2, 3, 4, 5}$. 
In Fig.~\ref{HRresonance}(b) and (c), we show the in-phase and quadrature components of the lock-in output for $R_1$ and $R_2$ resonances.
Note that the two resonances have $180^{\circ}$ opposite phases. 
This phase inversion is originated from the difference in the mechanisms of the $R_1$ and other resonance modes.
Since this difference is essential for the precise determination of the superfluid density and its critical exponent, we will discuss it in detail in the next Section.
There were many other resonance modes at higher frequencies than 2000 Hz, but they were not examined in detail in this work. 

We have measured the resonance curves such as those shown in Fig.~\ref{HRresonance}(b) and (c) at fixed temperatures between 0.7 and 2.4 K. 
The temperature dependencies of the superfluid density $\rho_{\mathrm s}$ and dissipation $Q^{-1}$ are obtained by fitting Lorenzian formula to the resonance curve at each temperature. 

\section{Results}\label{Results}
\subsection{Characteristics of resonance modes}\label{Modes}
The five resonances are characterized by two types of fundamental modes:  
The resonance modes $R_{2-5}$ are observed in the entire experimental temperature range $0.7 < T < 2.4 $ K. 
On the other hand, the lowest frequency mode $R_1$ was observed only below the superfluid transition temperature of $^4$He inside the Gelsil pores, $T_{\mathrm c}$. 
Moreover, the $R_1$ mode has an opposite phase relation, between the drive and output voltage, to the other modes as shown in Fig.~\ref{HRresonance}(b) and (c).  
These facts indicate that $R_1$ and $R_2$ are the fundamental modes of the resonator, while $R_{3-5}$ are higher order modes of the $R_2$ mode. % from that of the $R_{2-5}$ modes. 
%Hence we firstly describe the characteristics the $R_2$ modes that was studied in most detail among the $R_{2-5}$ modes. %and $R_1$ modes, both of which are useful to reveal the superfluid properties of helium in Gelsil. 

\subsubsection{The $R_2$ modes}\label{R2mode}
\begin{figure}[tb]
\centering
\includegraphics[width=1.0\linewidth]{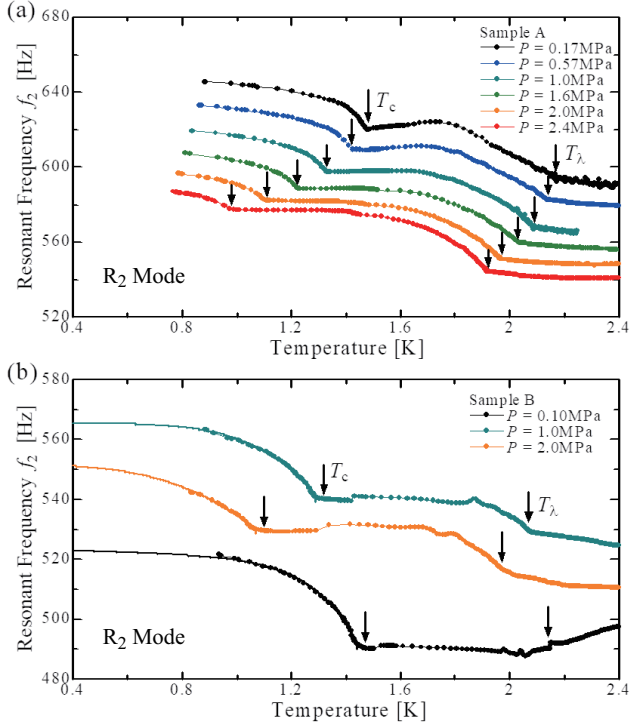}
\caption{(Color online) \label{freq_f_2}The resonant frequency of the $R_2$ mode, $f_{2}$, for different pressures as a function of temperature. 
Data are for (a) sample A and (b) sample B, respectively. 
Superfluid transition temperatures in bulk helium and in helium in Gelsil are indicated by $T_{\rm\lambda}$ and $T_{\rm c}$, respectively. 
Colored solid lines in (b) are the results of fitting to the temperature dependence of the superfluid density, $\rho_{\mathrm s}(T)$, obtained from the $f_1$ data shown in Fig. \ref{freq_f_1} using Eq. (\ref{f1}). 
The fitting is performed by multiplying a constant to $\rho_{\mathrm s}(T)$. }
\end{figure}

Figure \ref{freq_f_2} shows the temperature dependencies of $f_2$ for sample A and B at various pressures. 
The accompanied dissipation $Q^{-1}$ is discussed in the later section. 
The $R_2$ resonance was detected in the entire experimental temperature range $0.7 < T < 2.4 $ K.
As $T$ decreases from 2.4 K, $f_2$ gently increases (except for the data at $P=0.1$ MPa of sample B), and shows a sharp increase at temperatures that coincide with bulk $T_{\lambda}$. 
Below $T_{\lambda}$, $f_2$ shows a rounded temperature dependence, and then changes little around 1.6 K.
At a certain temperature denoted as $T_{\mathrm c}$, $f_2$ increases again and then saturates at lower temperatures. 
Since $T_{\mathrm c}$ is close to the superfluid transition temperature observed in the previous TO study\cite{YamamotoPRL2004}, it can be tentatively identified as the superfluid transition temperature of $^4$He in the present Gelsil samples.
In later section we determine $T_{\mathrm c}$ as a fitting parameter in the power-law fitting of the superfluid density in Gelsil.
The temperature dependence of $f_2$ is common in the data obtained at all pressures.
As the pressure increases, the overall behaviors shift to lower frequencies due to the increase in the mass of liquid $^4$He inside the resonator. 
In the measurement of sample B (Fig. \ref{freq_f_2} (b)), $f_2$ at 0.10 MPa is exceptionally lower than the data at higher pressures. 
We attribute this behavior to an accidental change of the tension of detector diaphragm $\sigma_{2}$ when the pressure increased from 0.1 to 1.0 MPa. 
This irregularity in the pressure dependence of $f_2$ does not affect the overall temperature dependence. 

\subsubsection{The $R_1$ mode}\label{R1mode}
In the measurement of sample A, it was difficult to determine the temperature dependence of resonant frequency $f_{1}$ due to the noise in the measurement electronics, which was particularly prominent at low frequencies at which the $R_1$ mode exists.
We then improved the electronics setup so that $f_{1}$ and dissipation were successfully determined for sample B.

\begin{figure}[tb]
\centering
\includegraphics[width=1.0\linewidth]{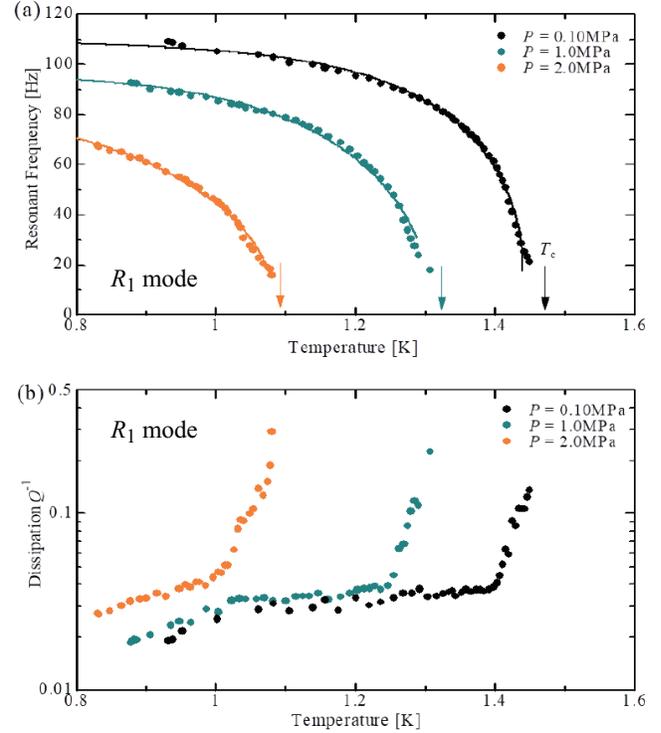}
\caption{(Color online) \label{freq_f_1} The lowest frequency mode $R_1$. 
(a) Resonant frequency $f_1$ at $P =$ 0.1, 1.0 and 2.0 MPa for sample B. 
$T_{\rm c}$, which is indicated by an arrow, is obtained from the power law fitting to $f_2^{-1}$ (see text). 
The location of $T_{\rm c}$ suggests that $f_1$ tends to zero at $T_{\rm c}$. 
Solid lines are the results of fitting using an arbitrary function in order to obtain the function $f_1(T, P)$.
(b) Corresponding dissipation $Q^{-1}$ obtained from the linewidth of the resonance curve. 
The divergent behavior suggests the absence of the resonance above $T_{\mathrm c}$. 
}
\end{figure}

The temperature dependencies of $f_{1}$ and dissipation $Q^{-1}$ are shown in Fig.~\ref{freq_f_1}.
Contrary to the measurements of the $R_{2-5}$ modes, the $R_1$ mode was identified only below a certain temperature near $T_{\mathrm c}$.
As $T$ increases from the lowest temperature,  $f_1$ monotonously decreases and tends to 0 Hz near $T_{\mathrm c}$. 
Simultaneously, $Q^{-1}$ increases in the vicinity of $T_{\mathrm c}$ with a divergent behavior, meaning that the $R_1$ resonance vanishes at $T_{\mathrm c}$.
%This divergent behavior of $Q^{-1}$ and the decrease in $f_1$ to zero indicate that the $R_1$ mode exists

\subsubsection{Identification of the resonance modes}\label{Identification}
Above the bulk $T_{\lambda}$, $^4$He is a normal viscous liquid with shear viscosity $\eta \sim 3 \times 10^{-6}$ Pa$\cdot$s. 
This gives the viscous penetration depth $\delta = \sqrt{2\eta/\rho \omega} \sim 3 \mu$m, where $\rho$ is the liquid density and $\omega$ the angular frequency of oscillatory flow. 
As $\delta$ is much larger than the pore diameter $d_{\mathrm p}$, the motion of the normal liquid $^4$He is tightly blocked in the nanopores of Gelsil, so that the liquid in RI is mechanically \textit{isolated} from the liquid in RO.
Therefore, the resonance modes observed above $T_{\lambda}$, i.e. the $R_{2-5}$ modes, can be modeled as a coupled oscillation among two diaphragms D1 and D2, which have tensions $\sigma_1$ and $\sigma_2$, respectively, and liquid $^4$He enclosed in RI.

\begin{figure}[tb]
\centering
\includegraphics[width=1.0\linewidth]{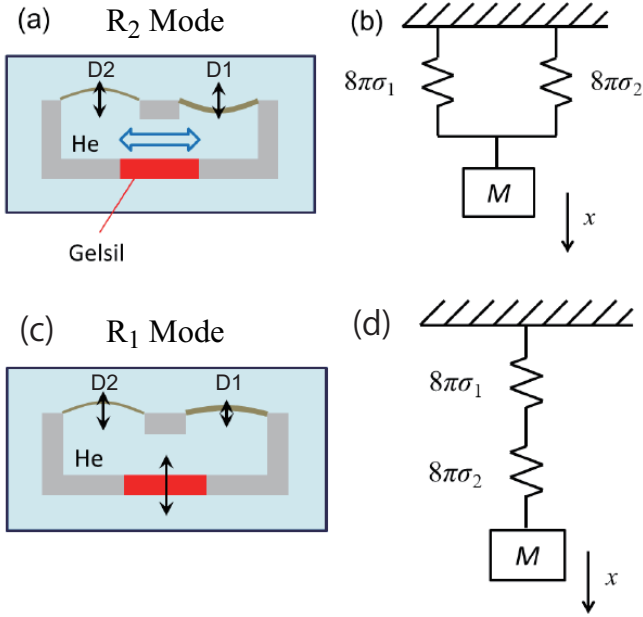}
\caption{(Color online) \label{f2f1mode} Schematic illustration of the two modes $R_{2}$ and $R_{1}$.
(a) In the $R_{2}$ modes, the detection diaphragm D2 oscillates \textit{in phase} with the drive diaphragm D1. 
(b) The mechanical analogue of the $R_{2}$ modes is expressed by a hydrodynamic mass $M$ hung under two parallel springs with spring constants $8\pi\sigma_1$ and $8\pi\sigma_2$ for D1 and D2, respectively. 
Here $\sigma$ denotes the tension of diaphragm.
(c) The $R_1$ mode occurs only when liquid helium flows through Gelsil. 
In this situation, two diaphragms oscillate $180^{\circ}$ out of phase. 
(d) The mechanical analogue consists of the mass $M$ hung by two springs $8\pi\sigma_1$ and $8\pi\sigma_2$ in series.
}
\end{figure}

The resonance curves shown in Fig. \ref{HRresonance} shows that at the $R_2$ resonance the detector diaphragm D2 oscillates \textit{in phase} with the drive diaphragm D1, as shown in the illustration of Fig. \ref{f2f1mode}(a).
This means that helium in RI can be treated practically as an \textit{incompressible} liquid.
We can then apply a simplified model for the $R_2$ mode, which is shown in Fig. \ref{f2f1mode} (b). 
It consists of an effective mass $M$ of liquid helium in RI hung under two parallel springs with effective spring constants $8\pi\sigma_1$ and $8\pi\sigma_2$.
The helium mass $M$ is a complicated function of the real mass, viscosity of liquid, and the geometry of RI. 
The effective spring constant of liquid $^4$He $\kappa$ is estimated to be $\kappa = A^2 / \chi V \sim 2.5 \times 10^5$ N/m, where $A$ is the total area of the diaphragms, $\chi$ the compressibility of liquid $^4$He ($1.2 \times 10^{-7}$ Pa$^{-1}$, and $V$ is the volume of RI. 
The diaphragm tensions $\sigma_1$ and $\sigma_2$ are obtained to be 150 (150) and 70 (10) N/m for samples A (B), respectively, from changes in the capacitance between the diaphragms and the fixed electrodes when they are biased with several DC voltages.  
This practical incompressibility of liquid leads to the simple mass-spring model of Fig. \ref{f2f1mode} (b).

Below $T_{\lambda}$, the viscosity-free superfluid component makes the effective mass $M$ decrease, resulting in the increase in $f_2$. 
However, the liquid in RI is still isolated from that in RO, because the liquid $^4$He in Gelsil behaves as normal, although LBECs are formed in the nanopores. 
When the liquid in Gelsil undergoes the superfluid transition at $T_{\mathrm c}$, the liquid can flow through the nanopores of Gelsil. 
Although the mechanism of the reduction of $M$ by the superfluid component is complicated, it does not influence the analysis because $f_2$ tends to be constant just above $T_{\mathrm c}$ (around 1.5 K).  

To conclude, at $T > T_{\rm c}$, the displacement of D2 is equal to that of D1 and liquid $^4$He in RI, while once $^4$He in Gelsil undergoes the superfluid transition, the displacement of D2 becomes smaller than others due to the superflow through Gelsil. 
Then the resonant frequency $f_{2}$ below $T_{\mathrm c}$ is obtained by the model of oscillating liquid $^4$He with mass $M$ under the restoring force of parallel spring constants $8\pi\sigma_1$ and $8\pi\sigma_2$. 
Assuming that the flow velocity of superfluid $^4$He through Gelsil is lower than superfluid critical velocity, $f_{\mathrm {2}}$ below $T_{\rm c}$ is obtained as
\begin{equation}\label{f2}
  f_{\mathrm {2}} = \frac{1}{2\pi}\sqrt{\frac{8\pi \left( \sigma_{1}+\sigma_{2} \right)}{\rho V - \alpha\rho_{\rm s}}},
\end{equation}
where $\rho$ is density of bulk liquid $^4$He, $\rho_{\rm s}$ the superfluid density of $^4$He in Gelsil, $V$ the total volume of inner and outer liquid, and $\alpha$ a coefficient depending on configuration of Gelsil. 

Before considering the $R_1$ mode, we briefly note on the $R_{3-5}$ modes. 
The $R_{3-5}$ modes were observed at $0.7 < T <  2.4$ K, as in the case of the $R_2$ mode. 
The $R_3$ mode has the same phase relationship between D1 and D2 as that of the $R_2$ mode. 
However, the $R_4$ and $R_5$ modes have an opposite phase relationship to the $R_2$ mode. 
These behaviors are seen in Fig. \ref{HRresonance} (a). 
Although the $R_{3-5}$ modes can be regarded as higher order modes of the fundamental $R_2$ mode, they are yet to be identified accurately. 
The inversion of the phase in the $R_4$ and $R_5$ modes might be caused by the small but finite compressibility of liquid helium. 
%They may therefore be identified as a kind of the higher order modes of the fundamental $R_2$ mode. 
%However, it is difficult to elucidate the origin of the high frequency modes. 
In Appendix\ref{R3mode}, we show the characteristics of the $R_3$ mode and discuss the critical exponent derived from $f_3(T)$.  

%For sample B shown in Fig.~\ref{freq_f_2}(b), contrary to sample A, sharp increases in $f_{2}$ at $T_{\rm c}$ are slightly dulled in the same manner under various pressures. 
%If the superfluid velocity of liquid $^4$He through Gelsil exceed a superfluid critical velocity, an effective friction on superflow depending on driving force should smear the increase in resonant frequency. 
%We made sure that the $T$ dependence of $f_{2}$ was not different as the driving AC voltage was changed, meaning that the dulled increase of $f_{2}$ in the vicinity of $T_{\rm c}$ is not attributed to critical velocity but to $\rho_{\rm s}$ itself affected by, probably, some unknown details of Gelsil nanopore structure. 
%Despite this difference, the superfluid transition temperatures $T_{\rm c}$ well agrees with each other. 

Next, we consider the resonance process of the $R_1$ mode.  
The $R_1$ resonance is observed only below $T_{\mathrm c}$, and the phase between drive and detection in the $R_1$ resonance curve has an opposite relation to that of the $R_2$ mode.
The $R_1$ mode is therefore attributed to pure superfluid motion through Gelsil. % and the resonant frequency $f_{1}$ is explained by a hydrodynamic consideration. 
When the driver diaphragm D1 pulles liquid $^4$He inside RI as illustrated in Fig. \ref{f2f1mode} (c), some amount of superfluid flows from RO to RI via Gelsil. 
This superflow pulles up the detector diaphragm D2 as it is much more flexible than D1. 
This mechanism explains the fact that D1 and D2 oscillate \textit{out of phase} at $f_{1}$, contrary to the other modes. 
We calculate the temperature dependence of $f_1$ by considering the hydrodynamic properties of liquid $^4$He in bulk and in Gelsil.

The resonant frequency $f_{1}$ can be written by a hydrodynamic inductance $L$ and a capacitance $C$, defined by $L(dJ_{\rm s}/dt) = -\Delta\mu/m_{^4{\mathrm {He}}}$ and $J_{\rm s} = (C/m_{^4{\mathrm {He}}})(d\Delta\mu/dt)$, where $J_{\rm s}$ is the mass current of superflow, $m_{^4{\mathrm {He}}}$ the mass of $^4$He atom and $\Delta\mu$ the chemical potential difference across the nanopore channel of Gelsil, as $2\pi f_{1} = \sqrt{1 / LC}$. 
In the case of superflow oscillation by restoring force from the diaphragms, $L = 1/\rho_{\rm s}\beta$, 
where $\beta$ is a structual coefficient depending on the detail of the channel, 
and $C = \rho^{2}A_{\rm d}^{2}(1/\sigma_{1} + 1/\sigma_{2})/8\pi$, 
where $A_{\rm d}$ is the area of each diaphragm. 
Assuming that bulk liquid $^4$He in RI is incompressible, $f_{\mathrm {1}}$ is given by
\begin{equation}
  f_{\mathrm {1}} = \frac{1}{2\pi}\sqrt{\frac{\rho_{\rm s}}{\rho^{2}}\beta\frac{8\pi \left(1/\sigma_{1} + 1/\sigma_{2} \right)^{-1}}{A_{\rm d}^{2}}}.
\label{f1}
\end{equation}
If we assume that Gelsil consists of $N$ parallel straight flow channels, the superfluid density of liquid $^4$He in Gelsil is estimated to be about an order of magnituide smaller than the bulk one. 
Because $\sigma_{1}$, $\sigma_{2}$, $A_{\rm d}$ and $\beta$ are constant determined by the experimental setup and $\rho$ is regarded as constant in the current temperature range, $f_{\mathrm {1}}$ represents the temperature dependence of the superfuid density $\rho_{\rm s}$. 

Because $f_{1}$ approaches 0 Hz and the dissipation diverges at $T_{\rm c}$, the precise determination of $\rho_{\rm s}(T)$ from $f_{1}$ is difficult in the vicinity of $T_{\mathrm c}$.
However, the formula of $f_{1}$, Eq. (\ref{f1}), enables us to determine precisely $\rho_{\rm s}(T)$ in the overall temperature range. 
We compare the temperature dependence of $\rho_{\mathrm s}$ obtained from the $f_1$ data to $f_2$ by the following procedure. 
From the experimental data of $f_1$, we obtain an arbitrary functional form of $f_1(T)$ for three experimental pressures. 
$f_1(T)$ is then converted to $\rho_{\mathrm s}(T)$ using Eq. (\ref{f1}). 
Finally, $f_2$ is calculated by Eq. (\ref{f2}) in which a constant $\alpha$ is determined as a fitting parameter.  

The colored solid lines in Fig.~\ref{freq_f_2} (b) show such a calculated $f_2(T \le T_{\mathrm c})$. 
It is remarkable that the calculated $f_2(T)$ from $\rho_{\rm s}(T)$ obtained by Eq.~(\ref{f1}) using a single parameter $\alpha$ agrees exactly with the data of $f_2$. 
This agreement validates to obtain the superfluid critical exponent $\zeta$ from the $f_2$ data not only of sample B but A.

\subsection{Dissipation of the $R_2$ mode}\label{dissipationR2mode}
The data of dissipation $Q^{-1}$ provides another essential information for elucidating the mechanism of superfluid transition of $^4$He in Gelsil. 
Measurement of dissipation is generally influenced by mechanical stability of the resonator. 
Since the measurement employing sample A was suffered from some unstable behaviors, we have improved the setup such as vibrational isolation. 
Here we discuss the sample B measurement, in which we believe that clear temperature dependence of dissipation below $T_{\mathrm c}$ were obtained without suffering from instability problem. 

\begin{figure}[tb]
\centering
\includegraphics[width=0.9\linewidth]{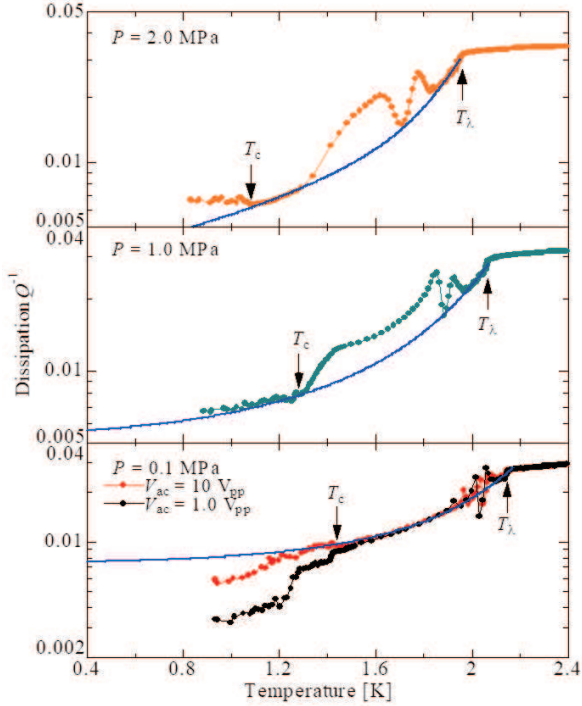}
\caption{(Color online) \label{dissipation_f_2} Temperature dependence of dissipation $Q^{-1}$ of the $R_2$ resonance in the measurement of sample B. 
From top to bottom panel, the pressure is 2.0, 1.0, and 0.10 MPa. 
$T_{\mathrm c}$ and $T_{\lambda}$ are indicated by arrows. 
Note that the dissipation is shown with different scales. 
Two data taken at different drive voltages $V_{\rm ac} = 1.0$ and 10.0 V$_{\rm {p-p}}$ are shown in the bottom panel ($P = 0.10$ MPa). 
Blue solid lines represent the possible contribution from bulk liquid $^4$He in the resonator (see text). 
}
\end{figure}

The temperature dependence of $Q^{-1}$ of the $R_2$ mode are shown in Fig.~\ref{dissipation_f_2}. 
We show two data sets for $P = 0.1$ MPa taken with two different drive voltages, $V_{\rm ac} = 1.0$ and 10.0 V$_{\rm {{p-p}}}$. 
Otherwise the data were taken at  $V_{\rm ac} = 10.0$ V$_{\rm {{p-p}}}$.
As $T$ decreases above $T_{\lambda}$, $Q^{-1}$ decreases gently. 
This is attributed to the temperature dependence of viscosity in bulk $^4$He. 
$Q^{-1}$ then falls sharply at $T_{\lambda}$, but at $T_{\lambda} > T > T_{\mathrm c}$ it decreases to about $1/6$ of the value at $T_{\lambda}$, with a seemingly nonsystematic change. 
We analyze the $Q^{-1}$ data at $T_{\mathrm c} < T < T_{\lambda}$ by assuming contributions from 
(1) change in viscosity of bulk liquid, 
(2) emergence of superfluid density in bulk liquid, and 
(3) coupling of the mechanical oscillation to second sound resonances. 
In order to analyze the effects of (1) and (2), we employ the data of the quartz tuning fork in the resonator cell. 
During the measurement, we have monitored the resonant frequency and dissipation of the fork at all the temperatures. 
The temperature dependencies of the frequency and the dissipation are quantitatively explained by taking into account the changes in the hydrodynamic mass of the fork and the viscous drag force in bulk $^4$He\cite{BlaauwgeersJLTP2007}. 

The dissipation of a fork in liquid $^4$He is given by $1/Af_{0} \propto \sqrt{\rho\eta}$, where $A$ is the amplitude of vibration, $f_{0}$ the resonant frequency and $\eta$ the effective viscosity of liquid. 
We monitored the pressure of the liquid in the volume RO, and found that the pressure changes with temperature according to the temperature dependence of liquid density. 
The actual change in $\rho\eta$ must include the effect of pressure change. 
Taking this effect into account, we determined the temperature dependence of $\rho\eta$ for three data sets ($P = 0.1, 1.0$ and 2.0 MPa) from the tuning fork measurement. 
The blue solid lines in Fig. \ref{dissipation_f_2} are $\sqrt{\rho\eta}$ for three pressures. 
They agree well with $Q^{-1}$ at temperatures just below $T_{\lambda}$ and near $T_{\mathrm c}$, but there are extra dissipation peaks in the intermediate temperature range. 
We attribute this excess dissipation to the coupling of the mechanical resonance $R_2$ to some standing wave modes of the superfluid second sound. 
We calculated the second-sound resonant frequencies of this particular resonator by the finite element method, and obtained a quantitative agreement with the temperatures of the dissipation peaks. 
Details of the analysis is given in Appendix \ref{Secondsound}.

Below $T_{\rm c}$, the dissipation $Q^{-1}$ taken at $V_{\rm ac} = 1.0$ V$_{\rm {p-p}}$ significantly decreases and deviates from the expected contribution of bulk liquid. 
This large reduction of dissipation is attributed to the emergence of superfluid mass current through Gelsil. 
The temperature dependence of $Q^{-1}$ should therefore be proportional to the superfluid density $\rho_{\rm s}$ of $^4$He in Gelsil. 
However, $Q^{-1}$ below $T_{\mathrm c}$ is not monotonous and a small peak-like structure is observed around 1.3 K. 
On the other hand, at $V_{\rm ac} = 10.0$ V$_{\rm {p-p}}$, $Q^{-1}$ changes little and starts to decrease at lower temperature.
In the data taken at $P = 2.0$ and 1.0 MPa, $Q^{-1}$ changes little or even slightly increases. 
These behaviors of $Q^{-1}$ suggest that an additional dissipation occurs and cancels the reduction of $Q^{-1}$ below $T_{\mathrm c}$.

The existence of excess dissipation in $^4$He in Gelsil is also supported by an experiment using porous Vycor glass, in which the confined $^4$He underwent superfluid transition at $T_{\rm c} =$ 1.97 K. 
At $T_{\rm c}$ the dissipation sharply decreased at any driving voltages, and no excess dissipation was observed. 
We conclude that the excess dissipation is unique to $^4$He in Gelsil and should be related to the mechanism of superfluid transition in much smaller pores than Vycor has.

\section{Discussion}\label{Discussion}
\subsection{Critical exponent of superfluid density $\rho_{\rm s}$}
In this section, we discuss the critical exponent of $\rho_{\rm s}$ derived from the $R_2$ mode in more detail than the previous publication\cite{TaniJPSJ2021}. 
Similar analysis and discussion for the $R_3$ mode in sample B are presented in Appendix\ref{R3mode}. 

In order to discuss the superfluid critical phenomenon in the vicinity of $T_{\mathrm c}$, we derive the temperature dependence of $\rho_{\mathrm s}$ of $^4$He in Gelsil using Eq.~(\ref{f2}) as the following formula,
\begin{equation}
\frac{1}{f_{2}^{2}(T_{\rm c},0)} - \frac{1}{f_{2}^{2}(T, \rho_{\rm s})} = \frac{\pi\alpha}{2 \left(\sigma_{1}+\sigma_{2} \right)}\rho_{\rm s}.
\label{rhos}
\end{equation}
Here the density of bulk liquid $^4$He is assumed to be independent of temperature in the range of interest ($0.7 < T < 1.6$ K). 
However, as seen in Fig.~\ref{freq_f_2}, the measured resonant frequencies just above $T_{\rm c}$ have a small but finite slope, in which the origin is unknown. 
We subtracted this as a background offset assuming a linear temperature dependence. 
Because $\sigma_{1}$, $\sigma_{2}$ and the structual coefficient $\alpha$ are constant, the left hand side of Eq.~(\ref{rhos}) determines precisely the temperature dependence of $\rho_{\rm s}$. 
%Similar analysis can be applied to the third resonsnce $f_{3}$. 
\begin{figure}[tb]
\centering
\includegraphics[width=1.0\linewidth]{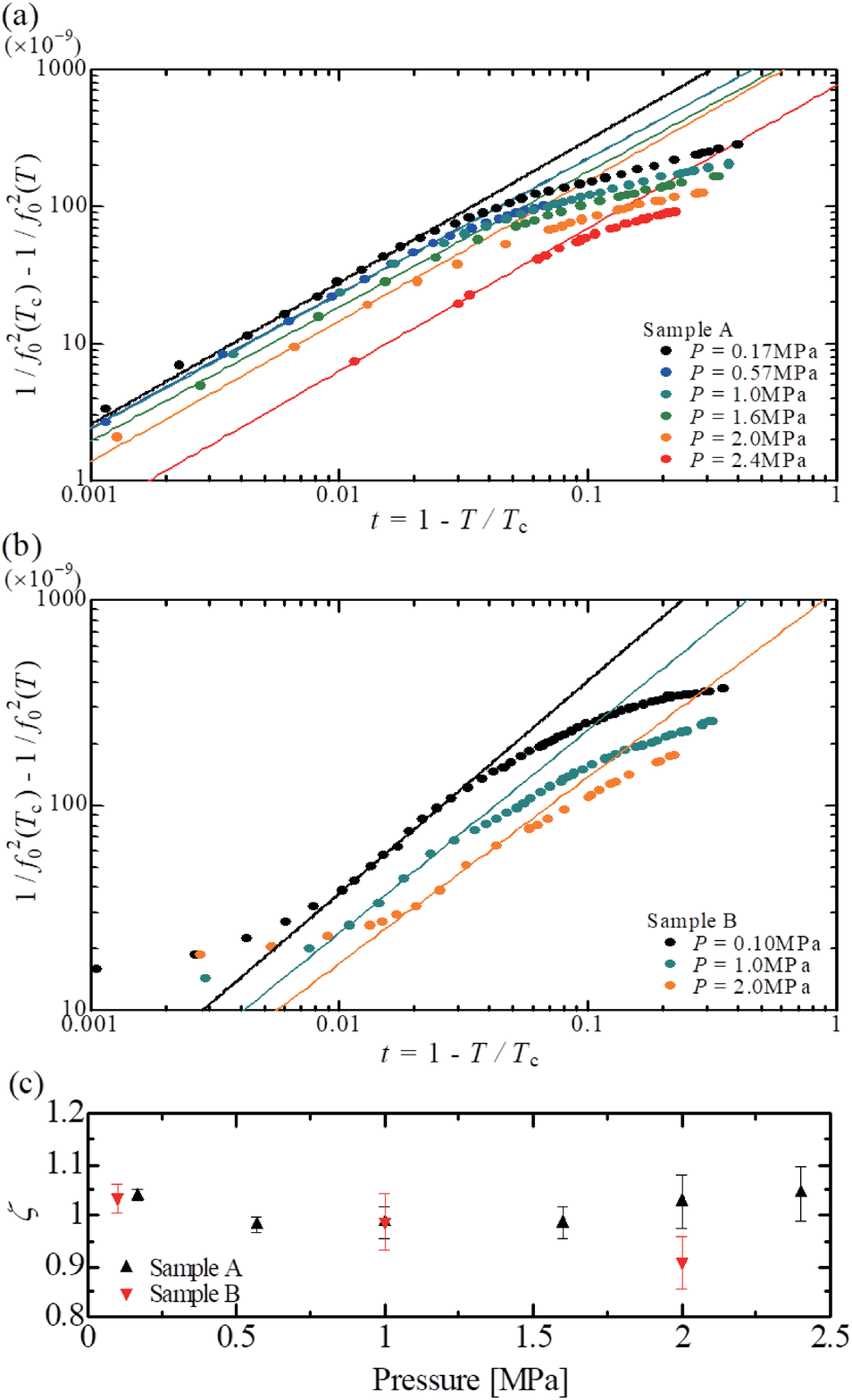}
\caption{(Color online) \label{CriticalExponent} Log-log plots of $1/f_{2}^{2}(T_{\rm c}) - 1/f_{2}^{2}(T)$, which is proportional to superfluid density $\rho_{\rm s}$, as a function of reduced temperature $t = 1 - T/T_{\rm c}$ under different pressures in (a) sample A and (b) sample B. (c) Critical exponent $\zeta$ versus pressure for each Gelsil sample.}
\end{figure}

In the discussion of critical phenomenon, the superfluid density $\rho_{\rm s}$ in the vicinity of $T_{\rm c}$ is written by a power law
\begin{equation}
\label{powerlaw}
  \rho_{\rm s} \propto \left| 1-T/T_{\rm c} \right|^{\zeta},
\end{equation}
where $\zeta$ is the critical exponent for superfluid density. 
In Fig.~\ref{CriticalExponent}, we show $\log$ - $\log$ plots of the quantity $1/f_{2}^{2}(T_{\rm c}) - 1/f_{2}^{2}(T)$, which is proportional to $\rho_{\rm s}$, obtained from the data shown in Fig.~\ref{freq_f_2}, as a function of reduced temperature $t = 1 - T/T_{\rm c}$. 
In these plots, we have determined $T_{\rm c}$ so as to make the longest straight lines in the $\log$ - $\log$ plots. 

All the data obey a power law in a limited range of $t$. 
The data of sample A shown in Fig.~\ref{CriticalExponent} (a) are well fitted with straight lines in the range $0.001 < t < 0.03$, while the plots for sample B shown in Fig.~\ref{CriticalExponent} (b) deviates from straight lines in the immediate vicinity of superfluid transition $t < 0.01$. 
This deviation reflects the result that $f_{2}$ near $T_{\mathrm c}$ of sample B is more rounded than that of sample A (see Fig.~\ref{freq_f_2}). 
Despite this smeared behavior in $\rho_{\rm s}$ in sample B, the $\log$ - $\log$ plots obey straight lines in the temperature range $0.008 < t < 0.04$. 
%Essentially, the critical phenomenon should be discussed in the vicinity of phase transition, and this exception of data in the immediate vicinity of $T_{\rm c}$ seems to be invalid. 
%However, such deviation of $\rho_{\rm s}$ in the immediate vicinity of $T_{\rm c}$ have been shown in other porous media, Vycor and Aerogel\cite{ChanPRL1988}. 
%Therefore we attributed such deviations in log-log plots to characteristics of porous media and they are essentially irrelevant to critical phenomenon. 
We discuss later the origin of the deviation from the power law.

The critical exponent $\zeta$ obtained from the slope of the straight lines in the $\log$ - $\log$ plots is shown in Fig.~\ref{CriticalExponent}(c). 
The error bars represent fitting errors in the plots. 
We found that $0.91 < \zeta < 1.04$ for both samples and at all pressures. 
No pressure dependence was observed. 
Although the data of sample B are smeared in the vicinity of $T_{\mathrm c}$, $\zeta$ are identical to the results of sample A, except for the case of $P = 2.0$ MPa, in which $\zeta = 1.03$ and 0.91 for sample A and B, respectively. 
Although the reason of the discrepancy is not clear, we note that the range of $t$ that shows the power law is rather narrow ($0.02 < t < 0.05$). 
\begin{table*}[tb]
 \caption{Superfluid critical exponents in various $^4$He systems}
 \label{Criticalexponenttable}
 \centering
  \begin{tabular}{cllllllll}
   \hline
   System & Pore size (nm) & Porosity (\%) & $P$ (MPa) & $T_{\mathrm c}$ (K) & $\zeta$ & Method & Ref. \\
  %  & & & $\times 10^8$ [m/sec] \\
   \hline \hline
   Bulk $^4$He & -- & -- & SVP & 2.172 & $0.6705 \pm 0.0006$ & Second sound & Goldner et al. \cite{GoldnerJLTP1993} \\
 Bulk $^4$He & -- & -- & SVP - 2.91 & 2.172 - 1.782 & 0.66 - 0.68 & Second sound & Greywall and Ahlers \cite{AhlersRMP1980}\\
 $n = 2$ vector  & \multirow{2}{*}{--} & \multirow{2}{*}{--} & \multirow{2}{*}{--} & \multirow{2}{*}{--} & \multirow{2}{*}{$0.67155 \pm 0.00027$} & \multirow{2}{*}{--} & \multirow{2}{*}{Campostrini et al.  \cite{CampostriniPRB2001}}\\ 
  (3DXY)  &   &   &   &   &   &   &  \\
   Vycor & 7 & 30 & SVP & 1.955 & $0.65 \pm 0.03$ & Fourth sound & Kiewiet et al. \cite{KiewietPRL1975} \\
   Porous Gold & 24, 75 & 58, 69 & SVP & 2.169 & $0.67 \pm 0.01$ & Torsional osc. & Yoon and Chan\cite{YoonPRL1997} \\
   Xerogel & 10 & 60 & SVP & 2.088 & $0.89 \pm 0.02$ & Torsional osc. & Chan et al. \cite{ChanPRL1988} \\
   Aerogel & not defined & 99.5 - 94 & SVP & 2.1717 - 2.1698 & $(0.72 - 0.81) \pm 0.01$ & Torsional osc. & Yoon et al. \cite{YoonPRL1998} \\
   Aerogel & not defined & 94 & SVP & 2.166 & $0.81 \pm 0.01$ & Torsional osc. & Wong et al. \cite{WongPRB1993} \\ 
   Aerogel & not defined & 93.6 & SVP - 2.9 & 3 mK below $T_{\lambda}$ & $0.755 \pm 0.003$ & Heat pulse & Mulders et al. \cite{MuldersPRL1991}   \\
   Gelsil & 3 & 50 & 0.1 - 2.4 & 1.47 - 1.0 & $1.0 \pm 0.1$ & AC flow & This work \\
   \hline
  \end{tabular}
\end{table*}

%As was discussed in the previous Letter\cite{TaniJPSJ2021}, 
The exponent $\zeta = 1$ is unprecedented. 
All the other confined $^4$He systems that were previously studied show $\zeta$ smaller than 1.  
We summarize the superfluid critical exponents in $^4$He systems in Table \ref{Criticalexponenttable}. 
%that $^4$He in Gelsil  of obtained critical exponents determined independently take values around 1, which indicates that $^4$He in Gelsil shows 4D quantum criticality at any finite temperatures. 
In bulk $^4$He, the critical exponent $\zeta_{\mathrm b}$ was determined to be $0.6705$ at SVP with great accuracy\cite{GoldnerJLTP1993}. 
Under pressure, detailed analyses taking confluent singularities into account concluded $0.66 < \zeta_{\rm b} < 0.68$\cite{AhlersRMP1980}. 
These values agree exactly with a critical exponent $-\nu$ for $n = 2$ vector (i.e. 3D XY) model calculated by Monte Carlo and high temperature expansion\cite{CampostriniPRB2001}.%\cite{LeGuillouPRL1977}.   
$\zeta$ of $^4$He confined in various porous materials had also been examined:  
In porous Vycor glass with 7 nm pore diameter and in two porous gold samples, $\zeta$ has been obtained to be 0.65 and 0.67\cite{KiewietPRL1975,YoonPRL1997}. It is remarkable that $\zeta$ is identical to the bulk one $\zeta_{\mathrm b}$ in these porous media. 
The bulk - like critical exponent may be a manifestation of the Harris criterion, in which, if the critical exponent of heat capacity, $\alpha$, is negative ($\alpha < 0$) in pure system, the critical phenomenon is not influenced by randomness or disorder\cite{HarrisJPC1974}. 
This is the case of superfluid $^4$He, because at the $\lambda$ transition $\alpha$ is obtained to be $-0.0127$\cite{LipaPRB2003}. 

On the other hand, $^4$He in aerogel shows interesting discrepancy\cite{ChanMuldersReppyPhysToday1996}. 
Aerogel consists of random network of silica strands with about 10 nm thickness and has large open volume. 
Thus the pore size is ill-defined and the porosity is extremely large (90 $\sim$ 99.5 \%) compared to other porous madia such as Vycor and Gelsil.
Torsional oscillator and heat capacity studies revealed that $^4$He shows a sharp superfluid transition, with $\zeta$ to be near 0.8, depending on the sample batch and porosity\cite{ChanPRL1988,WongPRB1993}. 
As the heat capacity peak in aerogel are strongly rounded, $\alpha$ was estimated by fitting the data to powerlaw in a finite temperature range, and was found to take a large negative value ($-0.9$) for low porosity samples. 
More recent study shows that both $\zeta$ and $\alpha$ approaches to the values of bulk $^4$He as the porosity increases from 95 to 99.5 \%\cite{YoonPRL1998}. 
To conclude, in quite contrast with the case of Vycor and porous gold, the Harris criterion does not hold in $^4$He in aerogel. 
$^4$He confined in xerogel, a nanoporous glass with the same structure as Vycor and Gelsil with pore size about 10 nm, also shows $\zeta = 0.89$, which does not agree with the Harris criterion\cite{ChanPRL1988,WongPRB1993}. 
Though it remains a matter of speculation, these discrepancies were attributed to correlation in disorder: i.e. The disorder in aerogel and xerogel are spatially \textit{correlated}, while Vycor and porous gold have uncorrelated disorder\cite{ChanMuldersReppyPhysToday1996,WongPRB1993}. 

The characteristics of the superfluid critical phenomenon in $^4$He in Gelsil differs from the abovementioned confined $^4$He systems in two observations,  
low transition temperatures that eventually reaches 0 K, and the critical exponent $\zeta$. 
$T_{\mathrm c}$'s of other systems are located at temperatures between 1.95 to 2.17 K at SVP, very close to the bulk transition $T_{\lambda}$, and do not exhibit any QPT at high pressures. 
The critical exponent $\zeta$ of $^4$He in Gelsil is larger than $\zeta$ in bulk and these porous materials. 
This implies that superfluid $^4$He in Gelsil belongs to a different universality class from in bulk or in other porous materials even at the highest transition temperature $T_{\rm c} =$ 1.48 K. 

In theory of critical phenomena, the superfluid order parameter $\Psi$ near $T_{\rm c}$ is expressed by $|\Psi| \propto (1 - T/T_{\rm c})^{\beta}$. 
The mean field theory gives the critical exponent $\beta = 0.5$\cite{NishimoriOrtiz}. 
As the superfluid density $\rho_{\rm s} =|\Psi|^{2}$, the critical exponent for $\rho_{\rm s}$ is $\zeta = 2\beta = 1$. 
Because the upper critical dimension of the XY model is 4, the 4D XY model should be described by the mean field theory. 
Fig.~\ref{CriticalExponent} (c) shows that the experimentally determined $\zeta$ is 1 at all pressure range we examined. 
We conclude that the superfluid transition in Gelsil belongs to 4D XY universality class even at finite temperatures far from 0 K. 

\subsection{4D XY criticality at finite temperatures}
\subsubsection{Absence of macroscopic superfluidity above $T_{\mathrm c}$}
As shown in Sec. \ref{R1mode}, the $R_1$ mode disappears at $T_{\mathrm c}$, i.e. $f_1$ tends to zero with diverging $Q^{-1}$.
This fact provides conclusive evidence that no macroscopic superfluidity exists throughout the Gelsil samples between $T_{\mathrm c}$ and $T_{\lambda}$. 
Although the absence of superfluidity between $T_{\mathrm c}$ and $T_\lambda$ was shown in the previous TO studies, the TO could not perfectly exclude the superfluidity of $^4$He in Gelsil above $T_{\mathrm c}$, as the TO is sensitive to the contribution from bulk $^4$He inside the bob\cite{YamamotoPRL2004}. 

\begin{figure}[tb]
\centering
\includegraphics[width=1.0\linewidth]{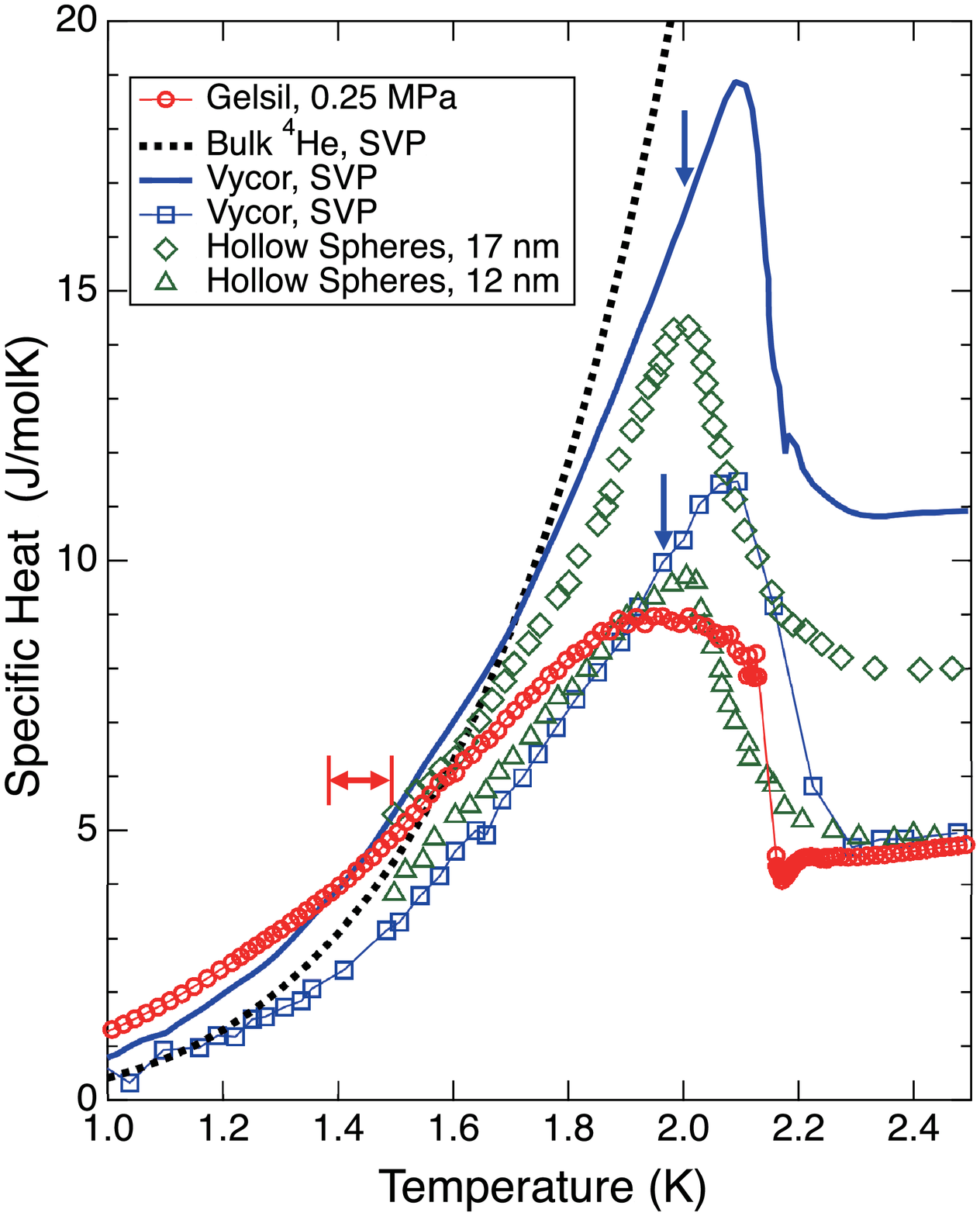}
\caption{(Color online) Specific heat of $^4$He in Gelsil and other systems. 
Red circles: $^4$He in Gelsil at $P = 0.25$ MPa\cite{YamamotoPRL2008}. Specific heat is derived from original heat capacity data by assuming $^4$He density to be that of bulk liquid. 
Red arrow indicates the onset temperature range of macroscopic superfluidity. 
Black dashed line: Bulk liquid at SVP, adopted from Ref. \cite{DonnellyBarenghi1998}. 
The specific heat at $T > T_{\lambda}$ is off the scale of the graph.
Blue solid line: $^4$He in Vycor at SVP measured by Zassenhaus\cite{Zassenhaus}. 
Blue arrow attached to the line indicates the onset temperature of superfluidity observed by torsional oscillator technique. 
Blue squares: $^4$He in Vycor at SVP measured by Brewer\cite{BrewerJLTP1970}. 
Blue arrow attached to a datum shows the onset temperature of superflow. 
Note that the two Vycor data sets have about 1.7 times difference in magnitude, and the superfluid onset temperatures are significantly lower than the temperatures of the specific heat peak. This is the same characteristics as that of $^4$He in Gelsil, in which the peak temperature is 1.95 K.
Green lozenges and triangles: $^4$He droplets formed in the nanosized spheric voids in Cu foils\cite{SyskakisPRL1985}.
17 and 12 nm indicate the mean diameters of the spheres, but liquid $^4$He form adsorbed films at the wall, i.e. in each void, liquid $^4$He forms a hollow sphere whose thickness is unknown.
\label{specificheat}
}
\end{figure}

Despite of the perfect absence of macroscopic superfluidity, the heat capacity shows a peak near the bulk $T_{\lambda}$, indicating a superfluid transition above $T_{\mathrm c}$\cite{YamamotoPRL2008}. 
Typical data of the specific heat are shown in Fig. \ref{specificheat} together with the data of other $^4$He systems. 
In this graph, the data of $^4$He in porous Vycor glasses\cite{Zassenhaus,BrewerJLTP1970} and of $^4$He droplets formed in microscopic bubbles in copper metal foils\cite{SyskakisPRL1985} are also plotted. 
The specific heat data of all the confined $^4$He systems have a blunted shape of the sharp $\lambda$-like peak of the bulk superfluid transition. 
Clearly, this "smeared-out" specific heat is caused by a finite-size effect in the superfluid transition\cite{GaspariniRMP2008}. 
We particularly take notice of the similarity between the specific heat of $^4$He in Gelsil and that of $^4$He droplets in copper\cite{SyskakisPRL1985}. 
The latter system consists of a number of independent hollow spheres of liquid $^4$He, in which liquid is formed on the metal wall with finite but unknown thickness, coexisting with helium gas inside the liquid sphere. 
The typical sphere sizes of the two data in Fig. \ref{specificheat} are 17 and 12 nm. 
It is remarkable that the data of $^4$He in Gelsil are qualitatively identical to the data of the 12 nm hollow spheres. 
This agreement leads us to conclude that, at the temperature of the specific heat peak, $T \sim 1.9$ K, $^4$He in Gelsil undergoes a finite-size superfluid transition without showing macroscopic superfluidity. 
(In this paper, we call the emergence of finite sized superfluid droplets a superfluid transition, although it is not a true phase transition in the thermodynamic limit.)
Then it is reasonable to speculate that the size of the superfluid droplets, i.e. LBECs, is limited by the pore size. 

\subsubsection{Superfluid transition and correlation length}\label{superfluidtransitioncorrelationlength}
\begin{figure*}[tb]
\centering
\includegraphics[width=1.0\linewidth]{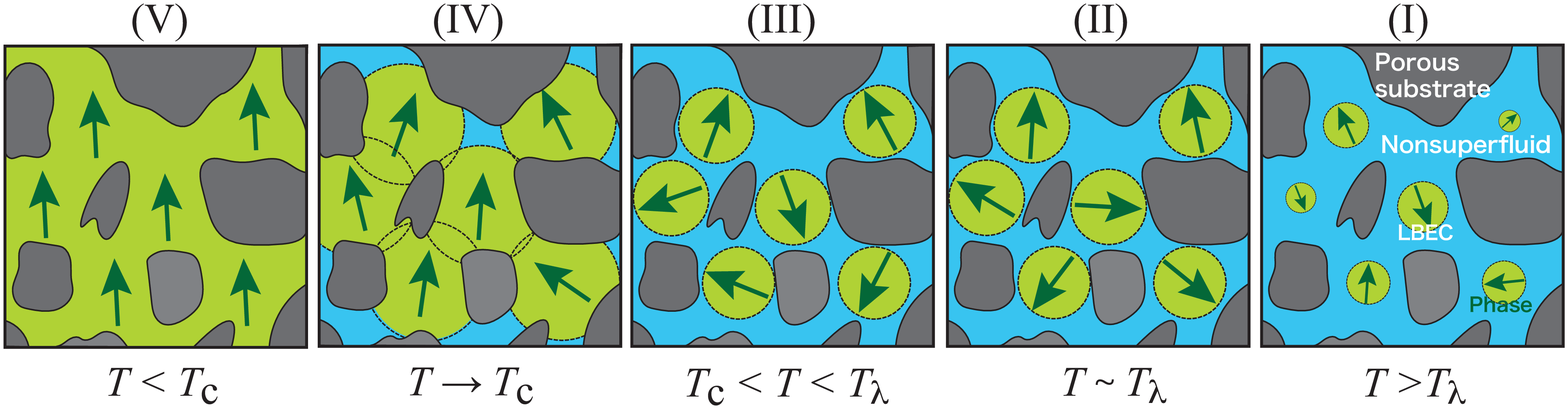}
\caption{(Color online) \label{LBEC} Illustration of the mechanism of superfluid transition of $^4$He in Gelsil. Phases of the superfluid order parameter are symbolically depicted by arrows. (I) At $T > T_{\lambda}$, seeds of LBECs start to grow, largely fluctuating in size and phase direction. 
(II) At $T $ slightly lower than the bulk $T_{\lambda}$, the growth of the LBECs is limited by a certain size $d$, which is a large part of the diameter of nanopores, resulting in the finite-size smearing of the specific heat peak. 
Each LBEC blob has an independent phase which fluctuates temporarily because of the suppression of atomic exchange in the narrow pore regions (indicated by blue color). 
(III) As $T$ decreases between $T_{\lambda}$ and $T_{\mathrm c}$, the LBEC blobs keep their sizes with fluctuating phase. 
(IV) As $T$ reaches to the vicinity of $T_{\mathrm c}$, the LBECs grow in size, and start to overlap. The phases among LBECs also start to be matched. 
(V) As $T$ passes $T_{\mathrm c}$, the phase becomes coherent throughout the Gelsil nanopores with a phase matching process with dissipation, resulting in superfluidity on macroscopic scale. 
}
\end{figure*}

\begin{figure}[tb]
\centering
\includegraphics[width=1.0\linewidth]{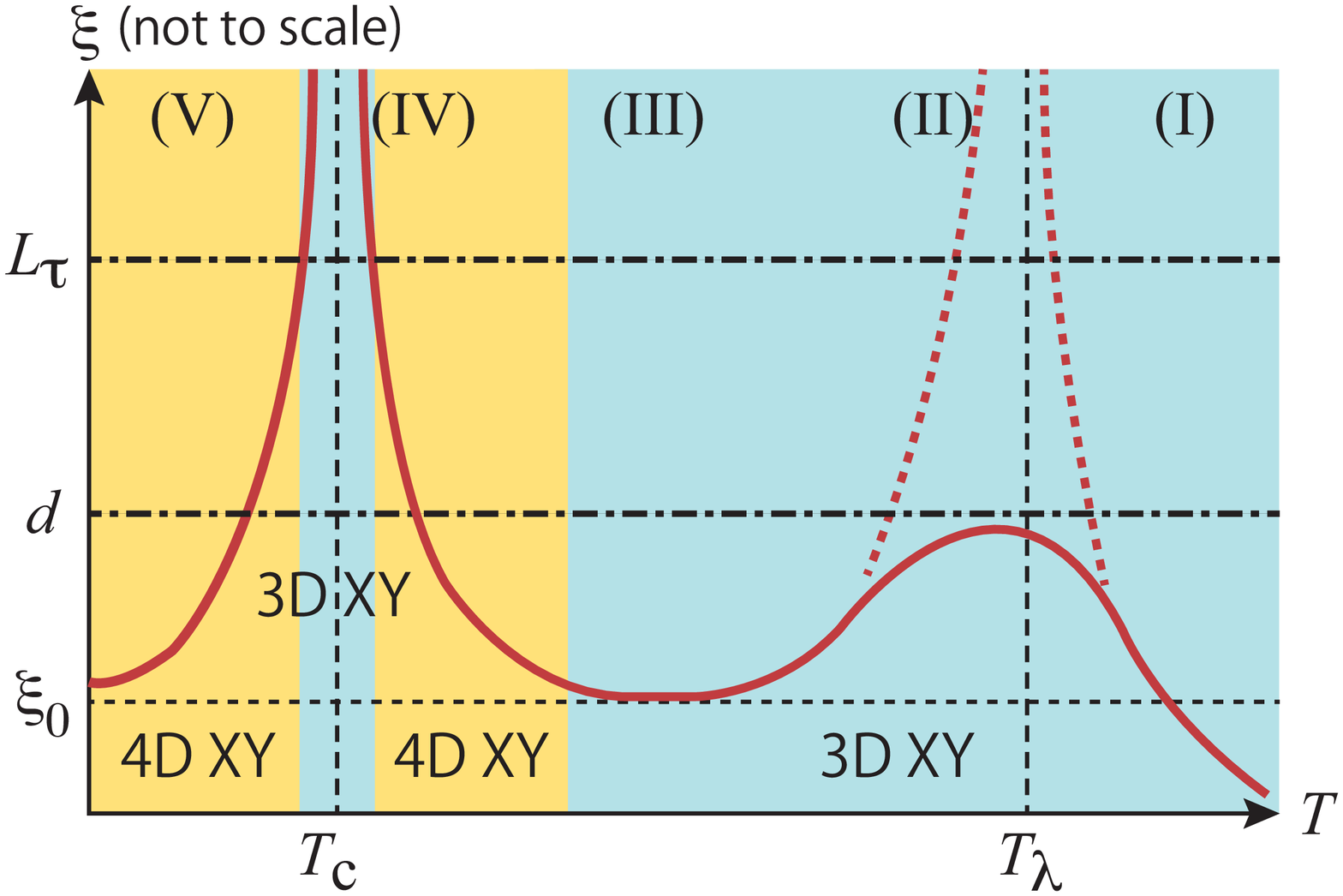}
\caption{(Color online) \label{correlationlength} Schematic illustration of the superfluid correlation (coherence) length $\xi(T)$ in $^4$He in Gelsil is shown by red solid lines. 
The figure is not drawn to scale. 
The regions shown by (I) -- (V) correspond to the illustrations in Fig. \ref{LBEC}. 
(I) As $T$ approaches $T_{\lambda}$, $\xi$ increases as in the case of bulk $^4$He. 
(II) The growth of $\xi$ is limited by a size $d$, which is possibly the major part in the pore size distribution of Gelsil. 
The dashes lines show $\xi$ in bulk $^4$He. 
(III) Below $T_{\lambda}$, $\xi$ decreases down to a length $\xi_0$. %, which is determined by the Josephson formula Eq. (\ref{Josephsonformula}). 
(IV) As $T$ approaches $T_{\mathrm c}$, $\xi$ increases with the 4D XY critical behavior. 
When the correlation length in the imaginary time dimension $\xi^z$, where $z$ is the dynamical critical exponent, reaches the size of quantum fluctuation $L_{\tau} = hc/k_{\mathrm B}T$, where $c$ is the the phonon velocity, a finite-size crossover from the 4D to the 3D XY criticality occurs, because the system size in the time dimension is limited by $L_{\tau}$. 
Then, $\xi$ diverges in 3D spatial dimensions at $T_{\mathrm c}$. 
(V) As $T$ passes $T_{\mathrm c}$, $\xi$ decreases, and the 4D XY criticality emerges again when $\xi^z < L_{\tau}$. 
At lower temperatures, $\xi$ decreases to $\xi_0$. 
}
\end{figure}
We propose a scenario of the formation of LBECs and the superfluid transition using illustrations in Fig. \ref{LBEC}, and the temperature dependence of the correlation length $\xi$ schematically shown in Fig. \ref{correlationlength}. 
The scenario described here is a corrected version of the scenario presented in the previous Letter\cite{TaniJPSJ2021}, in which the concept of the 4D-3D crossover was omitted. 
%The most important physics of $^4$He in nanoporous media is that the phase of the order parameter fluctuates by the suppression of positional exchange of helium atoms. 

At $T > T_{\lambda}$, indicated as the temperature region (I) in Figs. \ref{LBEC} and \ref{correlationlength}, seeds of LBEC are formed at the large spaces of the nanopores. 
The size of LBEC and the direction of the phase of the order parameter strongly fluctuate by thermal effect and suppression of the exchange of helium atoms. 
The correlation length $\xi(T)$, which is roughly the size of LBEC in this temperature regime, traces that of bulk helium (Fig. \ref{correlationlength}). 

As $T$ approaches the bulk $T_\lambda$ (the region (II)), the specific heat increases, but it is limited to a certain value at a temperature slightly below $T_\lambda$. 
This is attributed to the finite-size effect. 
The growth of the LBEC, i.e. the growth of $\xi$, is strongly limited at a length scale $d_{\mathrm p}$ determined by the nanoporous structure, because the correlation among LBECs is disturbed by the narrowness of the paths, in which the exchanges of helium atoms are strongly suppressed. 
The limit length $d_{\mathrm p}$ is probably the major part of the pore-size distribution shown in Fig. \ref{isotherms}, say, $3 < d_{\mathrm p} < 4$ nm.
The standstill of the growth of $\xi$ manifests itself in a rounded specific heat shown in Fig. \ref{specificheat}.
 
At $T$ between $T_\lambda$ and $T_{\mathrm c}$ (the region (III)), the specific heat decreases as in the case of bulk $^4$He. 
The specific heat of $^4$He in Gelsil below the peak temperature is explained by the contribution of rotons and phonons\cite{YamamotoPRL2008}. 
This fact strongly suggests that each LBEC might behave as a superfluid droplet, although the system does not exhibit macroscopic superfluidity; 
i.e. The fluctuation of the phase of the order parameter in each LBEC tends to vanish, but the phases among LBECs are uncorrelated. 
In the region (III), the correlation length will decrease to a constant $\xi_0$. 
Note that, in this temperature region, $\xi_0$ is no longer the size of LBEC, but is much smaller than it. 

In region (IV), as $T$ approaches $T_{\mathrm c}$, $\xi$ starts to increase in both the spatial (3D) and imaginary time (+1D) dimensions, and the LBECs start to overlap each other (Fig. \ref{LBEC}). 
Here we denote the correlation length in the imaginary time direction as $\xi^z$, where $z$ is the dynamical critical exponent, which is considered to be 1 in the present system\cite{SondhiRMP1997,EggelPRB2011,EggelThesis2011}. 
The increase in $\xi^z$ is limited by a length $L_\tau$. 
$L_\tau$ is given by the system temperature and the velocity of a collective excitation, which is typically the phonon (sound) velocity $c$\cite{EggelThesis2011}, 
\begin{equation}
\label{ltau}
L_\tau = hc/k_{\mathrm B}T,
\end{equation} 
where $h$ is Planck's constant and $k_{\mathrm B}$ is the Boltzmann constant.
For helium at $P = 0.1$ MPa and at $T = 1.5$ K, we obtain $L_\tau$ as 7.5 nm. 
This value is larger than the major part of the pore size, $2 \sim 5$ nm. 
When $\xi$ exceeds $L_\tau$, the critical behavior should show a crossover from 4D to 3D XY. 
The 3D correlation length $\xi$ diverges at $T_{\mathrm c}$, then start to decrease. 

Just below $T_{\mathrm c}$ (the region (V)), the phase coherence establishes, as illustrated in Fig. \ref{LBEC}. 
Again, the criticality shows a crossover from 3D to 4D XY, as $\xi^z$ becomes smaller than $L_\tau$. 
The 4D XY critical behavior in $\rho_{\mathrm s}(T)$ is observed in this regime. 
As $T$ decreases further, the fluctuation of the order parameter vanishes, so that $\xi$ tends again to the low temperature limit $\xi_0$.

The correlation length $\xi_0$ can be calculated by using the Josephson relation\cite{JosephsonPhysLett1966,FisherBarberJasnowPRA1973,ZassenhausPRL1999}
\begin{equation}\label{xi03D}
\xi_{(0)} = \frac{ k_{\mathrm B} T m^2_{\mathrm {^4He}} } { \hbar^2 \rho_{\mathrm {s(0)}} }.
\end{equation}
%where $m_{^4{\mathrm {He}}}$ is the mass of a $^4$He atom. 
Note that this formula is for the 3D case. 
Since this relation is believed to hold generally between the superfluid density and the correlation length, the index $0$ is parenthesized in $\xi$ and $\rho_{\mathrm s}$. 
The zero-temperature limit of the superfluid density $\rho_{\mathrm {s0}}$ is obtained as a coefficient of the powerlaw fitting Eq. (\ref{powerlaw}), 
$\rho_{\mathrm s} = \rho_{\mathrm {s0}}  \left| 1-T/T_{\rm c} \right|^{\zeta}$. 
From the fitting result for the $P = 0.1$ MPa data of sample B shown in Fig. \ref{CriticalExponent} (b), we obtain $\rho_{\mathrm {s0}}$ to be 249 kg/m$^3$. 
The corresponding $\xi_{0}$ is estimated to be 0.33 nm. 
This value is close to $\xi_0$ of bulk helium at SVP, 0.31 nm\cite{ZassenhausPRL1999}. 
However, $\xi_0$ should also be estimated in the 4D space using the dimensionality dependent Josephson formula. 
We discuss on the 4D correlation length in Appendix\ref{Josephsonrelation}. 
%Here we point out that $\xi_0$ estimated by several methods ranges from 0.33 to 2 nm, smaller than the pore size of Gelsil. 

\subsubsection{4D XY criticality at zero and finite temperatures}\label{4DXYcriticality}
The pressure dependence of zero-temperature superfluid density, $\rho_s(T = 0 {\mathrm K}) \propto (P_{\mathrm c} - P)$, and the $P - T$ phase boundary, $P_{\mathrm c}(0) - P_{\mathrm c}(T) \propto T^2$, were clearly explained as an emergence of 4D XY criticality by applying the disordered Bose - Hubbard model to the $^4$He in nanoporous structure\cite{EggelPRB2011,EggelThesis2011}. 
In this model, each site corresponds to a large part of pore that contains one LBEC, which consists of approximately 30 - 100 $^4$He atoms, and the sites are connected to neighbors by randam hopping matrix $J$. 
The $i$-th site has a random chemical potential $\mu_i$ and a random charging energy $V_i$, which is an increase in energy when a $^4$He atom is transferred to the LBEC at the $i$-th site from the nearest neighbor. 
It is thus an analogue of the charging energy of the superconducting Josephson junction.
$V_i$ is given by 
\begin{equation}
V_{i} = \left( \mathcal{V}_{i} \nu^2 \kappa \right) ^{-1},
\end{equation}
where $\mathcal{V}_{i}$ is volume of $i$-th site, $\nu$ the number density and $\kappa$ the compressibility of $^4$He\cite{EggelPRB2011,EggelThesis2011}. 
Using the bulk $^4$He compressibility to $\kappa$, and assuming that two atomic layers adjacent to the pore wall surface do not participate in superfluidity\cite{MakiuchiPRB2018}, $V_{i}$ is estimated to be 0.54 K. 
This estimation suggests that the superfluid transition occurring around 0.5 K is governed by quantum fluctuation;
i.e. The phase of a LBEC fluctuates by the suppression of change in the number of atoms in one site, which cost the energy $V_i$. 
This is the very origin of the emergence of 4D XY. 

This estimation of $V_i$ has an ambiguity in magnitude: In the previous work of Eggel et al.\cite{EggelPRB2011}, $V_i$ was estimated to be 1.4 K. 
The discrepancy between the present and previous estimations is caused by the discrepancies between the nominal pore sizes and the measured pore parameters of Gelsil samples. 
As shown in Fig. \ref{isotherms}, the two Gelsil samples of nominal pore sizes 2.5 and 3.0 nm have identical pore size distributions that are peaked at 3.0 or 3.8 nm. 
The distribution of the pore size makes the precise determination of $V_i$ difficult. 
Moreover, it has been suggested that, in general, the compressibility of liquid may be suppressed by confinement into nanoporous media due to the restriction of molecular motion\cite{GorJCP2015}. 
The suppression of $\kappa$ can also occur in the narrow pore regions of the Gelsil samples, resulting in a further increase in $V_{i}$. 
Therefore, we conclude that the quantum effect dominates the superfluid transition, even at $T_{\mathrm c} \sim 1.5$ K.   

The mechanism of the emergence of 4D XY universality class was discussed in the paper by Eggel et al.\cite{EggelPRB2011}. 
In the Bose - Hubbard model\cite{FisherPRB1989}, the 4D XY criticality is realized only in the case of absence of disorder and at the multicritical points in the $\mu$ - $J/V$ phase diagram (tips of the Mott lobes). 
In other superfluid - insulator phase boundaries than the tips of the lobes, the p-h symmetry is broken, so the system does not show the 4D XY criticality. 
In the $^4$He-Gelsil system, the 4D XY seemed unrealistic because it is difficult to tune the chemical potential. 
Moreover, in the presence of randomness, the p-h symmetry might also be broken because the number of helium atoms in each LBEC, $n_i$, is randomly distributed. 
Eggel et al. found, however, that the fluctuation in $n_i$ produces an exponential factor $e^{-2\pi^2 (\Delta n)^2}$ in a coefficient producing the p - h symmetry breaking in the Lagrangian density, where $\Delta n$ is the average of the deviation of particle number from the average. 
Since this factor is as small as $10^{-43}$, the p - h symmetry breaking becomes negligible. 
Next, the randomness could produce correlation in disorder in the imaginary time dimension (the correlated disorder\cite{WeinribHalperin}). 
This correlated disorder effect can alter the universality class from 4D XY to others, but such a crossover will appear only at the very vicinity of QCP, at $P \sim P_{\mathrm c}$ and near 0 K. 
When the system is away from the QCP, the effect of correlated disorder is negligible. 
As a result, the 4D XY criticality can be possessed at finite temperatures. 
We propose that the robustness of 4D XY criticality is sustained not only at 0 K but at finite temperatures, because the charging energy $V_i$ for LBEC is large in the nanopores of Gelsil samples. 
On the other hand, in $^4$He in other porous media such as Vycor, $V_i$ is estimated to be about 10 mK\cite{EggelPRB2011}. 
This small $V_i$ makes the critical phenomena in other porous materials completely classical, which, in the case of Vycor, is 3D XY\cite{KiewietPRL1975,ZassenhausPRL1999,Zassenhaus}.  

In the imaginary time dimension, % of imaginary time $\tau = \hbar / k_{\mathrm B}T$, 
the system size is limited by $L_\tau$ which was derived above. 
In analogy with the 3D - 2D finite - size crossover observed in superfluid $^4$He slabs\cite{GaspariniRMP2008}, a crossover from the 4D to 3D XY ciriticality is expected to occur when the correlation length $\xi^z$ in the time direction reaches $L_{\tau}$. 
Thus, the critical exponent $\zeta$ may change from 1 to other value such as 0.67, as in the case of porous Vycor and gold\cite{KiewietPRL1975,YoonPRL1997}. 
In the result of the power-law fitting shown in Fig. \ref{CriticalExponent}, an upward deviation from the $\zeta = 1$ line is observed in sample B. 
This might be an indication of the 4D - 3D crossover, but it is hard to explain the absence of such deviations observed in sample A. 
% the direction of the deviation is opposite to the downward shift expected in the 4D - 3D crossover, in which the critical exponent must change to a value smaller than 1, even if disorder affects the exponent from 0.67 as in the case of $^4$He in aerogel\cite{WongPRB1993,YoonPRL1998}. 
In the previous letter we have attributed the deviation from the $\zeta = 1$ powerlaw in sample B to the thickness of the glass sample (2 mm), which is twice the thickness of sample A\cite{TaniJPSJ2021}.  
These interpretations should be reconsidered with more detailed experimental study in the vicinity of $T_{\mathrm c}$.

We note that a similar theoretical interpretation, the 4D - 3D XY crossover, was proposed by Franz and Iyengar\cite{FranzPRL2006} to explain the $T$ - linear superfluid density and the square-root dependence of $T_{\mathrm c}$ on hole doping in highly underdoped YBCO superconductors\cite{BrounPRL2007}.

\subsection{The scaling of superfluid density}\label{Scaling}
Eggel proposed that the observed physical quantities are described by a universal scaling function that depends on the ratio of the extent in the imaginary time dimension and $\xi^z$\cite{EggelThesis2011}. 
%He demonstrated that the superfluid density $\rho_{\mathrm s}$ measured by the TO\cite{YamamotoPRL2004}  
Here we describe the scaling argument and the result of fitting using the data of $\rho_{\mathrm s}$ obtained for samples A and B. 

In the theory of finite-size scaling, the free energy density $f_{\mathrm s}$ is given by a scaling function $F$, 
% as The scaling function is obtained a singular part of the free energy $f_{\rm s}$, which is written by
\begin{equation}
 % f_{\rm s} \sim \xi^{-(d+z)} \sim \delta^{\nu((d+z))},
 f_{\mathrm s} \sim F\left( \delta, L \right) \sim \delta^{x_{f_{\mathrm s}}/x_\delta} F\left( \delta^{-1/x_\delta}/L\right).
\end{equation}
Here $\delta$ is a control parameter related to $\xi$ as $\delta \sim \xi^{-1/\nu}$, $x_{f_{\mathrm s}}$ and $x_\delta$ is the scaling dimensions of $f_{\mathrm s}$ and $\delta$, and $L$ is the length scale in real spaces. 
Since the critical exponent of the correlation length $\nu=1/x_\delta$, 
\begin{equation}
 % f_{\rm s} \sim \xi^{-(d+z)} \sim \delta^{\nu((d+z))},
 f_{\mathrm s} \sim \delta ^{\nu x_{f_{\mathrm s}}} F \left( \xi / L \right) .
\end{equation}
As the scaling dimension of $f_{\mathrm s}$ is $d$, $f_{\mathrm s}$ can be written as 
\begin{equation}
  f_{\rm s} \sim F \left( \delta, L, \beta \right) \sim \delta^{\nu(d+z)} F \left( \xi /L, \xi^{z} /\beta \right),
\label{scaling1}
\end{equation}
including the scaling in the imaginary time dimension $\hbar \beta = \hbar /k_{\mathrm B}T$, where $\beta$ is the inverse temperature.

Superfluid density $\rho_{\rm s}$ is defined by the helicity modulus\cite{FisherBarberJasnowPRA1973} derived from the free energy.
The change in free energy density $\Delta f$ caused by the spatial change (twist) in the phase of the order parameter $\Delta \phi$ is 
\begin{equation}
  \Delta f \sim \left( \Delta\phi /L \right)^2\rho_{\rm s}.
\label{scaling2}
\end{equation}
%where $\Delta f$ is change in free energy density accompanied with phase change $\Delta\phi$.
In order for $\Delta f$ to satisfy Eq.~(\ref{scaling1}) and Eq.~(\ref{scaling2}),
\begin{equation}
  \rho_{\rm s} \sim \xi^{2} \delta^{\nu(d+z)} \sim \xi^{2-d-z}.
\end{equation}
Assuming that the scaling function of the superfluid density, $R_{\rho_{\rm s}}$, is a function of the ratio of the correlation length $\xi_{\tau} \sim \xi^{z}$ to the scale $\beta$ in the temporal direction, the superfluid density is written by
\begin{equation}
  \rho_{\rm s} \sim R_{\rho_{\rm s}} \left( \frac{\xi^{z}}{\beta} \right) \xi^{2-d-z}.
\end{equation}

We recall that $\xi = \delta^{-\nu}$ and the control parameter of superfluid transition of $^4$He in Gelsil is pressure, i.e. $\delta = P$.
Using the reduced pressure $p = 1 - P/P_{\rm c}$, where $P_{\rm c}$ is the critical pressure at QCP, and the critical exponents in 4D XY, $\nu = 1/2$ and $\nu(d+z-2) = 1$, the scaling of superfluid density is described by
\begin{equation}
  \frac{\rho_{\rm s}}{p} \sim R_{\rho_{\rm s}} \left( T / \sqrt{p} \right).
\end{equation}

\begin{figure}[tb]
\centering
\includegraphics[width=0.9\linewidth]{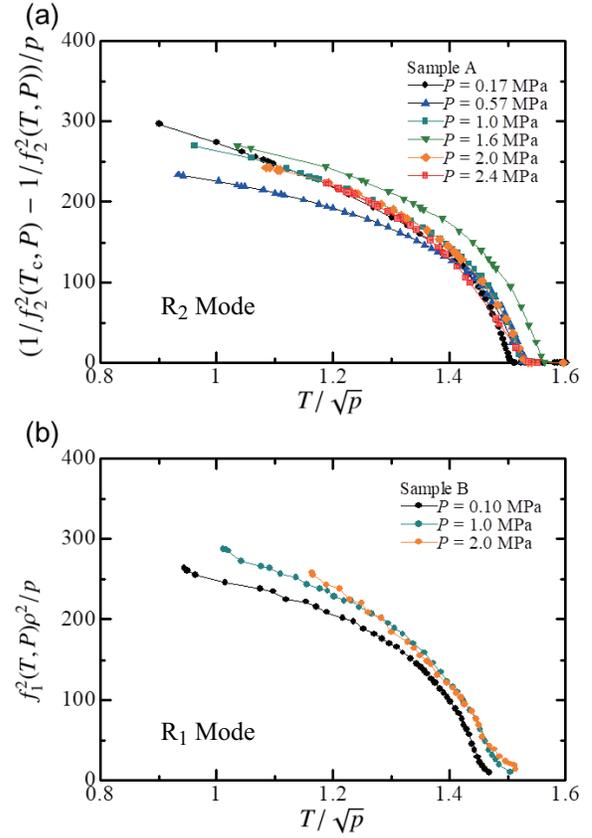}
\caption{(Color online) \label{scaling} Scaling of the superfluid density as a function of $T/\sqrt{P}$. Data are taken from (a) sample A, the $R_2$ mode, and (b) sample B, the $R_1$ mode. 
The quantities of vertical axes are proportional to $\rho_{\rm s} / P$.}
\end{figure}

The scaling of superfluid densities for each samples are shown in Fig.~\ref{scaling}. 
For sample A, the superfluid density $\rho_{\rm s}$ is evaluated by Eq. (\ref{rhos}) because the lowest mode $f_{1}$ was not precisely measured for sample A. 
For sample B, the pressure dependence of the resonant frequency $f_{2}$ was not monotonous due to the mechanical instability of D2 diaphragm. 
In the latter, therefore, the scaling of $\rho_{\rm s}$ is examined using Eq. (\ref{f1}). 
Both the quantities are proportional to $\rho_{\rm s}$, but $f_{1}^2$ is multiplied by $\rho^{2}$ to take the pressure dependence of density of bulk liquid $^4$He into account. 

As shown in Fig.~\ref{scaling}, most of the data sets successfully collapse onto a single curve, the scaling function. 
At some pressures, there exist deviations, which seem to be shift in temperature direction. 
There seems no monotonous deviations against pressure.
We attribute the deviations to unknown experimental errors such as the determination of pressure.
The scaling analysis is another strong evidence of the 4D XY criticality.

\subsection{Dissipation anomaly by phase alignment}%\label{dissipationanomaly}
The transition from the LBEC states to the macroscopic long-range ordered state is a process that  the phases of LBECs which initially point to random directions are aligned as the system is cooled passing $T_{\mathrm c}$, as shown in Fig. \ref{LBEC} (IV) and (V). 
During this phase alignment process, excess dissipation should be produced, and it can be detected as a change in the linewidth of the superfluid resonances. 
We calculate the excess dissipation using the Josephson - Anderson equation, which was proposed to discuss the decay of superflow by phase slippage\cite{JosephsonPhysLett1962,AndersonRMP1966,PackardRMP1998}.
Here we analyze the data of sample B, in which the dissipation was stably measured during the experimental runs. 

%\subsubsection{Background dissipation}
As we pointed out in Sec. \ref{dissipationR2mode}, in the dissipation $Q^{-1}$ of the superfluid resonator, there are contributions originated from the bulk liquid ((1) viscosity, (2) superfluid density, and (3) coupling to the second sound). 
The dissipation other than the second sound effect, i.e. the change in $\sqrt{\rho\eta}$, is calculated as the blue solid lines in Fig. \ref{dissipation_f_2}. 
The calculated curves basically agree with the temperature dependence of $Q^{-1}$ between $T_{\lambda}$ and $T_{\mathrm c}$, and some structures are attributed to the the coupling of second sound modes to the $R_2$ resonance. 
The coupling to the second sound is discussed later in this section and in Appendix \ref{Secondsound}.

\begin{figure}[tb]
\centering
\includegraphics[width=1.0\linewidth]{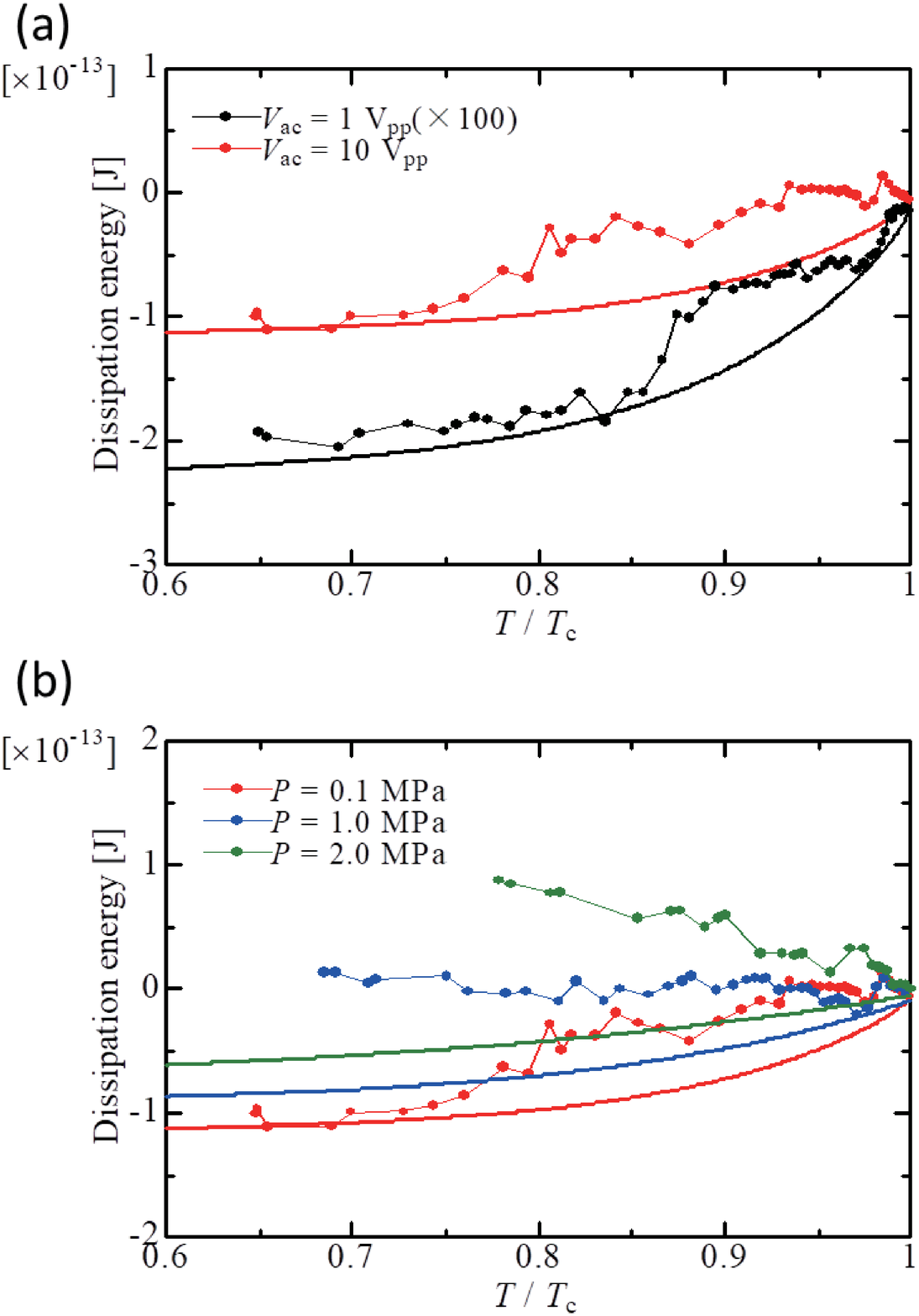}
\caption{(Color online) \label{dissipationbelowTc} Temperature dependence of dissipation energy in the $R_2$ mode of the sample B resonator, after subtracting the bulk liquid contribution $\sqrt{\rho\eta}$. 
 (a) Black and red data are taken at driving voltage $V_{\rm ac} =$1 V$_{\rm {p-p}}$ and 10 V$_{\rm {p-p}}$, respectively. The liquid pressure is 0.1 MPa. 
 The 1 V$_{\mathrm {{p-p}}}$ data are 100 times enlarged in the energy scale. 
 (b) Data taken at three pressures. Drive is 10 V$_{\mathrm {{p-p}}}$. 
 All solid lines are calculated negative contribution in the bulk liquid caused by the superflow between RO and RI through Gelsil. 
 Line colors show the voltages and pressures of the same colored data points. 
%Solid lines are estimated dissipation reduced by the superflow using $\rho_{\rm s}$ obtained by Eq.~(\ref{f1}).
}
\end{figure}

The dissipation energy after which the part of $\sqrt{\rho\eta}$ is subtracted is shown in Fig. \ref{dissipationbelowTc}. 
Here, in order to discuss the energy scale, $Q^{-1}$ is converted into the dissipation energy $\Delta E$ by the definition $Q = 2\pi E/\Delta E$, where $E$ is the total energy stored in the system. 
$E$ equals to the work by electric AC force, estimated by
\begin{equation}
  E = \int_{0}^{T} \frac{V_{\rm b}V_{\rm ac}}{\epsilon A_{\rm d}} C_{1}^{2} 2\pi f_{2} x_{0} \cos^{2}{\left( 2\pi f_{2} t \right)} dt,
\end{equation}
where $T$ is the period of the resonance, $x_{0}$ the amplitude of the displacement of driver diaphragm D1 and $C_{1}$ the capacitance between D1 and E1. 

The temperature dependence of the dissipation energy below $T_{\rm c}$ consists of two contributions. 
One is the negative contribution by the occurrence of superflow through Gelsil. 
% and the other is an additional dissipation characteristic to Gelsil. 
Below $T_{\mathrm c}$, the bulk liquid in RI hydrodynamically connects to the liquid in RO.
The mass superflow through the Gelsil nanopores results in the reduction in $\sqrt{\rho\eta}$. 
The reduction of dissipation induced by the superflow will depend on the details the flow structure in Gelsil (e.g. tortuosity of the flow paths) and the structure of the resonator. 
Although it is difficult to evaluate quantitatively these structural effects, the dissipation reduction will be simply proportional to the superflow mass. 
The superflow mass is determined by the superfluid density in Gelsil, $\rho_{\rm s}$, the superfluid velocity $v_{\mathrm s}$ inside the nanopores and the structure of the nanopores. 
Assuming that $v_{\mathrm s}$ has no temperature dependence, the negative dissipation is determined by the temperature dependence of $\rho_{\rm s}$. 
In fact, the data of $P = 0.1$ MPa taken at 1 V$_{\rm {p-p}}$ (black dots in Fig. \ref{dissipationbelowTc} (a)) agrees well with the curve obtained from the temperature dependence $\rho_{\rm s}$ (black line in the same Figure). 
We have therefore adjusted a coefficient to match the $T$ dependence of $\rho_{\rm s}$ obtained from the $R_1$ mode Eq.~(\ref{f1}) with the data at low temperatures. 
For the data of $P = 0.1$ MPa taken at 10 V$_{\rm {p-p}}$, the negative contribution to the dissipation was calculated in the same manner.
At pressures 1.0 and 2.0 MPa, the negative terms were determined with reference to the data of 0.1MPa at 10V$_{\rm {p-p}}$. 
These calculations are shown by solid lines in Fig.~\ref{dissipationbelowTc} (b). 
\begin{figure}[tb]
\centering
\includegraphics[width=0.95\linewidth]{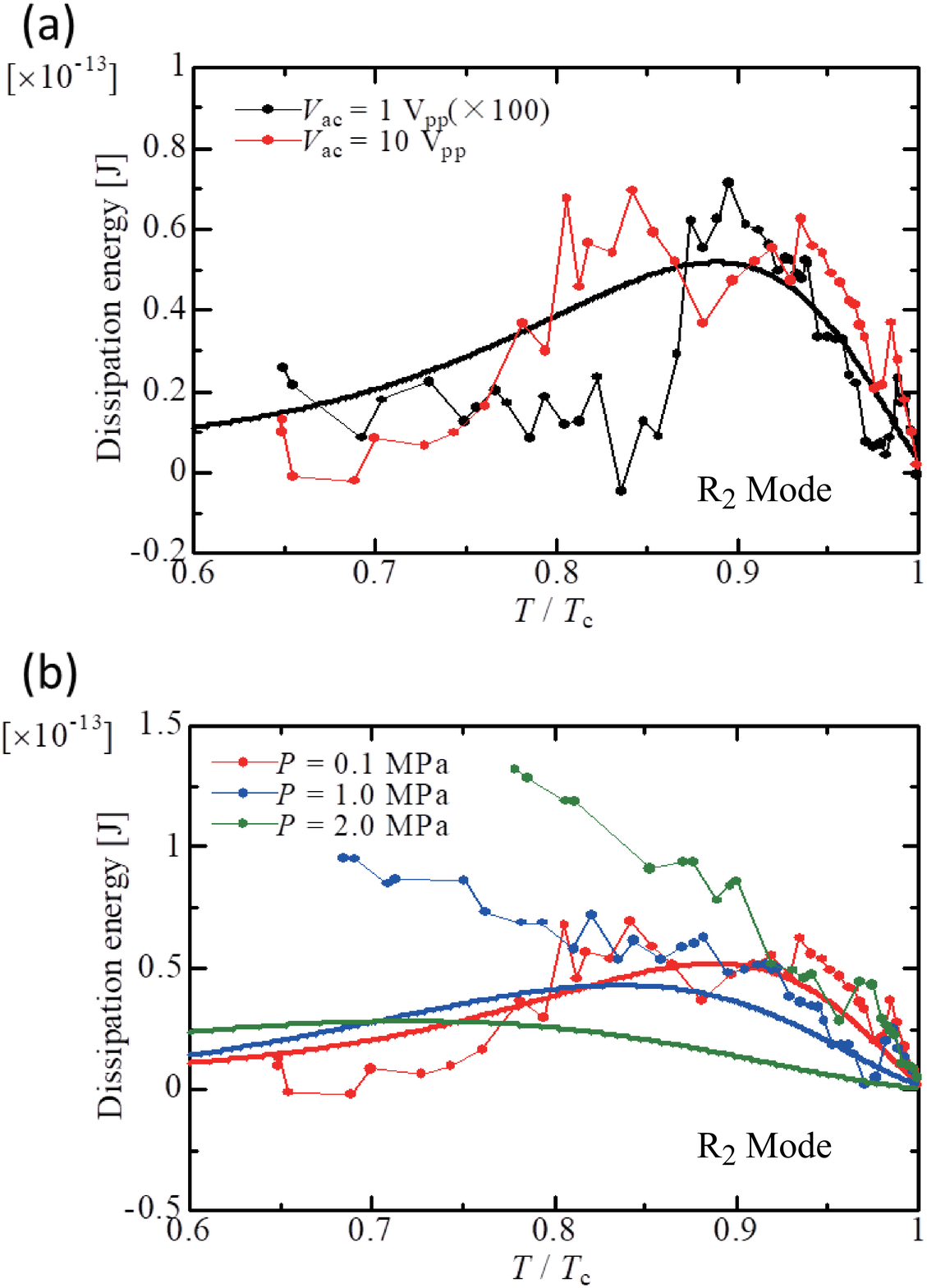}
\caption{(Color online) \label{dissipationanomaly} 
Temperature dependence of the dissipation anomaly. 
All the data correspond to the data in Fig. \ref{dissipationbelowTc} (a) and (b). 
(a) $P = 0.1$ MPa. Black and red data are taken at driving voltage $V_{\rm ac} =$1 V$_{\rm {p-p}}$ and 10 V$_{\rm {p-p}}$, respectively. The liquid pressure is 0.1 MPa. 
 The 1 V$_{\mathrm {{p-p}}}$ data are 100 times enlarged in the energy scale. 
 (b) Data taken at pressures 0.1, 1.0 and 2.0 MPa, and at 10 V$_{\mathrm {{p-p}}}$ drive. 
 All solid lines represent calculated dissipation energy using Eq.~(\ref{da}). 
 Line colors show the voltages and pressures of the same colored data points. 
}
\end{figure}

Subtracting the negative dissipation contribution from the data below $T_{\rm c}$, we find excess dissipation anomalies. 
The excess dissipation is shown in Fig.~\ref{dissipationanomaly}. 
Although the data are scattered, a broad peak or an increase is observed.  
In the data taken at $P = 0.1$ MPa, the dissipation has a peak at around $0.8 < T/T_{\mathrm c} < 0.9$. 
On the other hand, at pressures 1.0 and 2.0 MPa, the excess dissipation increases with decreasing temperature. 
We find that, in the 0.1 MPa data shown in Fig.~\ref{dissipationanomaly} (a), the height of the peak at 10 V$_{\rm {p-p}}$ is about 100 times the height of the 1 V$_{\rm {p-p}}$ data.
This means that the observed excess dissipation depends on the driving force. % and on a slope of $\rho_{\rm s}(T)$. 

We propose a model to explain the excess dissipation as an anomaly originated from the the alignment event of the mismatched phases among LBECs. %participating in the macroscopic superfluid in Gelsil and another one newly connected to the superflud. 
%At $T > T_{\mathrm c}$, each LBEC is isolated as a nanoscale superfluid, and the phase angle of the order parameter is not necessarily synchronized each other. 
As the LBECs overlap near or at $T_{\mathrm c}$, the phases of many LBECs start be matched each other with fluctuation. % he phase coherence is achieved by frequent exchanges of $^4$He atoms between LBECs, which is supported by the energy of mechanical resonance, leading to emergence of the additional dissipation on the resonance. 
Under the periodic driving force applied from the bulk liquid outside the Gelsil sample, the phase matching and mismatching events occur repeatedly during measurements over many periods of oscillation. This produces a statistic dissipation event which can be detected by the resonator response. 

In the Josephson - Anderson formula\cite{JosephsonPhysLett1962,AndersonRMP1966,PackardRMP1998}, changes in the phase angle of the superfluid order parameter produce the changes in energy, which is written by
\begin{equation}
  \frac{\partial\Delta\phi}{\partial t} = - \frac{\Delta\mu}{\hbar},
\end{equation}
where $\Delta\phi$ and $\Delta\mu$ are differences in superfluid phase and chemical potential between two spatially separated arbitrary points, respectively. 
In the present experimental situation, we take these differences between a macroscopically grown Bose-Einstein condensate producing macroscopic superflow via Gelsil and a LBEC droplet which participate newly to the condensate. 
When the phase of the newly connected LBEC is changed to the phase of the macroscopic superfluid, the phase of the LBEC tends to coincide with that of the macroscopic condensate, i.e. $\Delta\phi \rightarrow 0$, within a characteristic time scale $\tau$. 
The change of the chemical potential difference, i.e. energy cost per $^4$He atom, is given by $\Delta\mu = \hbar\Delta\phi / \tau$. 
Therefore, the change in energy per one additional LBEC is 
\begin{equation}
  \tilde{n}\Delta\mu = \tilde{n}\frac{\hbar\Delta\phi}{\tau}, 
\end{equation}
where $\tilde{n}$ is the mean condensate number per LBEC. 
$\tilde{n}$ is given by $\tilde{n} = \rho_{\rm L} \mathcal{V} / m_{^4{\mathrm {He}}}$, where $\rho_{\rm L}$ is the condensate density of LBEC, $\mathcal{V}$ the mean volume of LBEC. % and $m_{^4{\mathrm {He}}}$ the mass of $^4$He atom. 
Because LBECs grow from $T_{\rm\lambda}$, which is sufficiently higher than $T_{\rm c}$, $\rho_{\rm L}$ should be different from the macroscopic superflid density $\rho_{\rm s}$.
The superfluid density $\rho_{\rm s}$ is determined by the number of LBECs participating in the macroscopic superfluid with the local condensate density $\rho_{\rm L}$.

The phase coherence is achieved by positional exchanges of helium atoms between LBECs, which are carried by the flow induced by the resonator.
Assuming that the flow rate between LBECs develops toward low temperature proportionally to the macroscopic superfluid density $\rho_{\rm s}(T)$, and that the density carried between completely linked LBECs is $\rho_{\rm L}$, the mass current density between LBECs is $(\rho_{\rm L} / \rho_{\rm s}(0)) \rho_{\rm s}(T) v_{\rm s}$. 
The phase coherence is established when the mass carried from another LBEC spatially separated by a distance $l$ reaches the mass in a LBEC of density $\rho_{\rm L}$. 
Then, the characteristic time scale $\tau$ of phase matching event is
\begin{equation}
  \frac{1}{\tau} = \frac{1}{\rho_{\rm L}} \frac{\rho_{\rm L}}{\rho_{\rm s}\left(0\right)} \rho_{\rm s}\left(T\right) \frac{v_{\rm s}}{l} = \frac{\rho_{\rm s}\left(T\right)}{\rho_{\rm s}\left(0\right)} \frac{v_{\rm s}}{l}
\end{equation}

The dissipation event is caused by a newly connected LBEC at a given temperature. 
On the other hand, the LBECs that already participate to the macroscopic superfluid at sufficiently high temperature have phase coherence, and suffer no change in their phase because of the much larger energy cost than that of the small LBEC case. 
The number of LBECs contributing to the dissipation event, $\Delta N(T)$, is the number of the LBECs newly participating in the macroscopic superfluid within the temperature range $T \sim T+\Delta T$. 
Because the rate of LBEC belonging to the macroscopic superfluid equals to the rate $\rho_{\rm s}(T) / \rho_{\rm L}$, 
\begin{equation}
  \Delta N(T) = \frac{N_{0}}{\rho_{\rm L}}\left(\frac{\partial}{\partial T}\rho_{\rm s}(T)\right)\Delta T,
\end{equation}
where $N_{0}$ is the total number of LBECs. 
The characteristic temperature scale $\Delta T$ should correspond to the energy scale which makes the phase of a LBEC decoupled from the macroscopic superfluid. 
We assume that this energy equals to the energy of phase matching event, $\Delta T = \Delta\mu / k_{\rm B}$. 
Thus, the total dissipation energy by resolving the phase mismatch is
\begin{align}
  \Delta E &= \tilde{n}\Delta\mu \Delta N \nonumber \\
  &= \frac{\mathcal V}{m_{^4{\mathrm {He}}}k_{\rm B}}\left(\frac{\hbar\Delta\phi}{l}\right)^{2}\frac{N_{0}}{\rho_{\rm s}^{2}(0)}\left(\frac{\partial}{\partial T}\rho_{\rm s}\left(T\right)\right) \rho_{\rm s}^{2}\left(T\right) v_{\rm s}^{2}
\label{da}
\end{align}
Because superfluid phase has a degree of freedom $\phi = \phi + 2\pi m$ ($m$ is an integer), the phase matching can progress in either increasing or decreasing direction, which is decided by the energy cost.
Therefore, the maximum phase difference is effectively $\pi$ and $\Delta \phi$ equals to $\pi /2$ as a mean value. 

The dissipation energies calculated from Eq.~(\ref{da}) are shown in Fig.~\ref{dissipationanomaly} for different pressures. 
The calculated dissipation at $P = 0.1$ MPa has an excellent agreement with the data without any fitting parameters. 
The data at 1.0 MPa also agrees with the calculated line except for low temperatures, and the data at 2.0 MPa is significantly larger than the line at all temperature range. 
The data for 0.1 MPa taken at 1 V$_{\rm {p-p}}$ agrees with the calculated dissipation, in which the factor 100 is multiplied. 
This agreement ensures the dependence on the superfluid velocity $v_{\rm s}$ in Eq.~(\ref{da}) on the assumption that $v_{\rm s}$ is propotional to the drive voltage $V_{\rm ac}$.

The disagreement between the calculation and experimental data at high pressures, 1.0 and 2.0 MPa, may be originated from the coupling of the mechanical resonance to some standing wave modes of the superfluid second sound. 
We discuss the coupling of the second sound in Appendix \ref{Secondsound}. 
The coupling may be the origin of the additional dissipation structures and some anomalies in $f_2$ at $T_{\lambda} > T > T_{\mathrm c}$, which are shown in Fig. \ref{dissipation_f_2} and Fig. \ref{freq_f_2}, respectively. 
At $P = 1.0$ and 2.0 MPa, the frequencies of some standing wave modes are close to $f_2$ at temperatures below $T_{\rm c}$, and the modes have little temperature dependence because of the second sound velocity has a minimum. 
The coupling of the $R_2$ mode to the second sound modes with little temperature dependence may result in an additional dissipation in a broad range below  $T_{\rm c}$, 
This speculation is consistent with the fact that at 0.1MPa the dissipation data agree well with the calculation of the phase matching - induced dissipation. 
At 0.1 MPa, the sound velocity is so large that all the modes have much higher frequencies than $f_2$. 
To conclude, the excess dissipation below $T_{\mathrm c}$ is quantitatively accounted for the mechanism of phase alignment during the growth of macroscopic superfluid.  

\subsection{The phase diagram}
\begin{figure}[tb]
\centering
\includegraphics[width=0.8\linewidth]{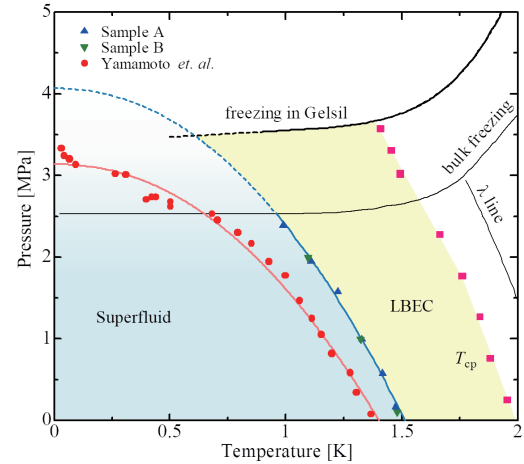}
\caption{(Color online) \label{phasediagram} The $P -T$ phase diagram. 
$T_{\mathrm c}$'s are plotted for the Gelsil samples A and B as blue and green triangles, respectively. 
Red points show $T_{\rm c}$ determined by the previous torsional oscillator study for a Gelsil sample of different batch\cite{YamamotoPRL2004}. 
Results of fittings to power law $P_{\rm c}(0) - P_{\rm c}(T) \propto T^{2.13}$ are indicated as blue and red lines\cite{EggelPRB2011,EggelThesis2011}. 
The temperatures of the peak in heat capacity, $T_{\mathrm {cp}}$, are shown by red squares\cite{YamamotoPRL2008}. 
Black lines show the bulk freezing curve and the superfluid $\lambda$ line.
}
\end{figure}
The $P$-$T$ phase diagram obtained in this and previous works is summarized in Fig.~\ref{phasediagram}.  
In this work, the data of phase boundary was obtained only at $P < 2.5$ MPa, over which bulk $^4$He freezes in the resorviors.
$T_{\rm c}$ is 1.45 K at 0.1 MPa, and decreases, as $P$ increases, down to 0.98 K at 2.4 MPa. 

It is remarkable that $T_{\mathrm c}(P)$ shows no difference between the results of sample A and B, whereas in the sample used in the previous TO study $T_{\mathrm c}$ was significantly lower than the present samples. 
The difference of $T_{\mathrm c}$ between the present samples A and B and the previous one is 0.08 K at 0.1 MPa, but it increases up to 0.25 K at 2.4 MPa. 
Kobayashi et al. also found that $T_{\mathrm c}(P)$ depends on the time of annealing of the glass sample for dehydration\cite{KobayashiJPSJ2010}. 
Even in a single sample, longer annealing time lowered $T_{\mathrm c}$ at high pressure region, although $T_{\mathrm c}$ near 0 MPa showed no change for the additional annealing. 
From these observations, the factors that determine $T_{\mathrm c}(P)$ is yet to be elucidated, although $T_{\mathrm c}(P)$ seems very sensitive to the nanoporous structure. 
In our scenario of the superfluid transition, it is essential that the LBECs are connected each other via narrow paths in which the positional exchange of helium atoms are strongly suppressed. 
This may be the origin of the sample - dependent $T_{\mathrm c}(P)$, in particular at high pressure regimes ($P > 2$ MPa). 
It is plausible that, in the narrow bottleneck paths (say, 2 nm) between LBECs, $^4$He atoms are hard to move particularly at high pressures, in which the interatomic correlation is enhanced. 

As in the previous work\cite{EggelPRB2011}, the phase boundary is fitted by $P_{\rm c}(0) - P_{\rm c}(T) \propto T^{2.13}$ with better quality than the fitting to the previous TO data. 
The quantum critical pressure $P_{\rm c}(0)$ is obtained for samples A and B to be 4.08 MPa. 
This is higher than the freezing pressure of $^4$He in Gelsil which was determined in a previous measurement, $P_{\rm f} \sim 3.7$ MPa\cite{YamamotoJPSJ2007}. 
Although the freezing pressure of $^4$He in Gelsil is unknown in the present Gelsil samples, the QPT may be masked by the solid phase. 
Such a masking, even if it exists, will not influence the quantum critical nature observed at lower pressure.

\section{Conclusion}
We have precisely determined the critical exponent of superfluid density of $^4$He confined in two nanoporous Gelsil glass samples by a newly developed mechanical resonance technique. 
This work has revealed a number of features that are essential for elucidating the mechanism of the superfluid transition. 
The superfluid critical exponent is obtained as $\zeta = 1.0 \pm 0.1$ for two glass samples under all pressure ranges where the experiment was performed.
This means that superfluid $^4$He in Gelsil exhibits a 4D XY quantum criticality even at finite temperatures, contrary to general expectation for quantum critical phenomena, in which the phase transitions at any finite temperatures should be classical.
This apparent contradiction is explained by a hypothesis about the mechanism of superfluid transition that the macroscopic superfluidity emerges by the sequential growth of local coherence between nearest neighbor LBECs, where the correlation length is still finite. 
This proposed mechanism has also been confirmed in the context of the evaluation of dissipation that is expected to occur accompanied with the superfluid transition, as the energy cost for resolving the phase mismatch between LBECs. 
In this work, a novel mechanism of the superfluid transition of $^4$He in nanoporous media has been definitely established.  
This outcome will be a key for further understanding general confined $^4$He systems. 

%We revealed that the superfluid transition of $^4$He in Gelsil exhibits 4D XY quantum criticality at phase transition of any finite temperatures, which is an unique nature to superfluid $^4$He in Gelsil. 
To our knowledge, superfluid $^4$He in Gelsil is the first example of bosonic 4D XY system in real matter.
Further studies of QPT in superfluid $^4$He in nanoporous media is expected to stimulate theoretical studies of 4D XY in the Bose-Hubbard model\cite{ProsniakSR2019} and 4D XY QPTs proposed in fermionic systems\cite{BrounPRL2007,FranzPRL2006}.

\section*{Acknowledgments}
We are grateful to Kazuyuki Matsumoto and Tomoki Minoguchi for fruitful discussions. 

\appendix

\section{The $R_3$ mode}\label{R3mode}
\begin{figure}[tb]
\centering
\includegraphics[width=0.95\linewidth]{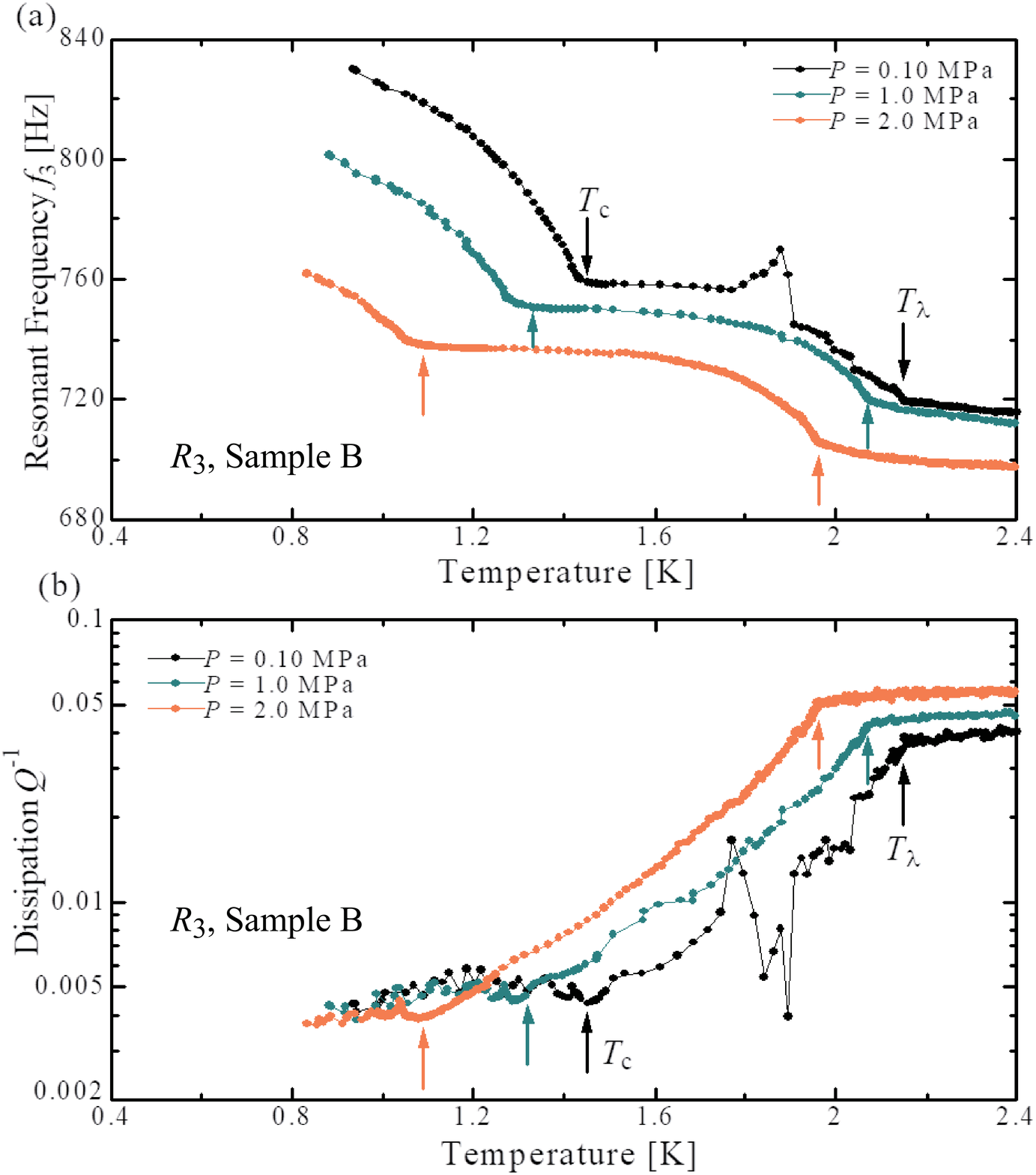}
\caption{(Color online)\label{R3FreqQinv}
The $R_3$ mode in the sample B resonator. Temperature dependencies of (a) resonant frequency $f_3$ and (b) dissipation  $Q^{-1}$ at $P = 0.10$, 1.0, and 1.0 MPa. 
Arrows indicate $T_{\lambda}$ and $T_{\mathrm c}$. 
}
\end{figure}

We have made a measurement of temperature dependence of the $R_3$ mode for sample B. 
The data of $f_3$ and $Q^{-1}$ are shown in Fig. \ref{R3FreqQinv}. 
%We see that the overall temperature dependence of $f_3$ is similar to that of $f_2$: 
As the temperature decreases, $f_3$ shows a sharp increase at bulk $T_{\lambda}$, then saturates below 1.6 K, 
and shows a steep increase at $T_{\mathrm c}$ inside Gelsil.
$Q^{-1}$ decreass at $T_{\lambda}$ and decreases monotonously, except for large scatter in the data of $P = 0.1$ MPa at $1.8 < T < 2.1$ K. 
The origin of the large scatter of $Q^{-1}$ and the accompanied jumps in $f_3$ is unknown. 
At $T_{\mathrm c}$, $Q^{-1}$ turns to increase and shows a broad peak. 

The temperature and pressure dependencies of the $R_3$ mode are similar to those observed in the $R_2$ mode. 
The $R_3$ mode can therefore be a higher order mode of $R_2$, but the detail of the higher order mode has not yet been elucidated. 
As an attempt, we apply Eq. (\ref{f2}), the formula of $f_2$ obtained from the spring-mass model in Fig. \ref{f2f1mode} (b), to the temperature depenence of $f_3$ below $T_{\mathrm c}$. 

\begin{figure}[tb]
\centering
\includegraphics[width=0.95\linewidth]{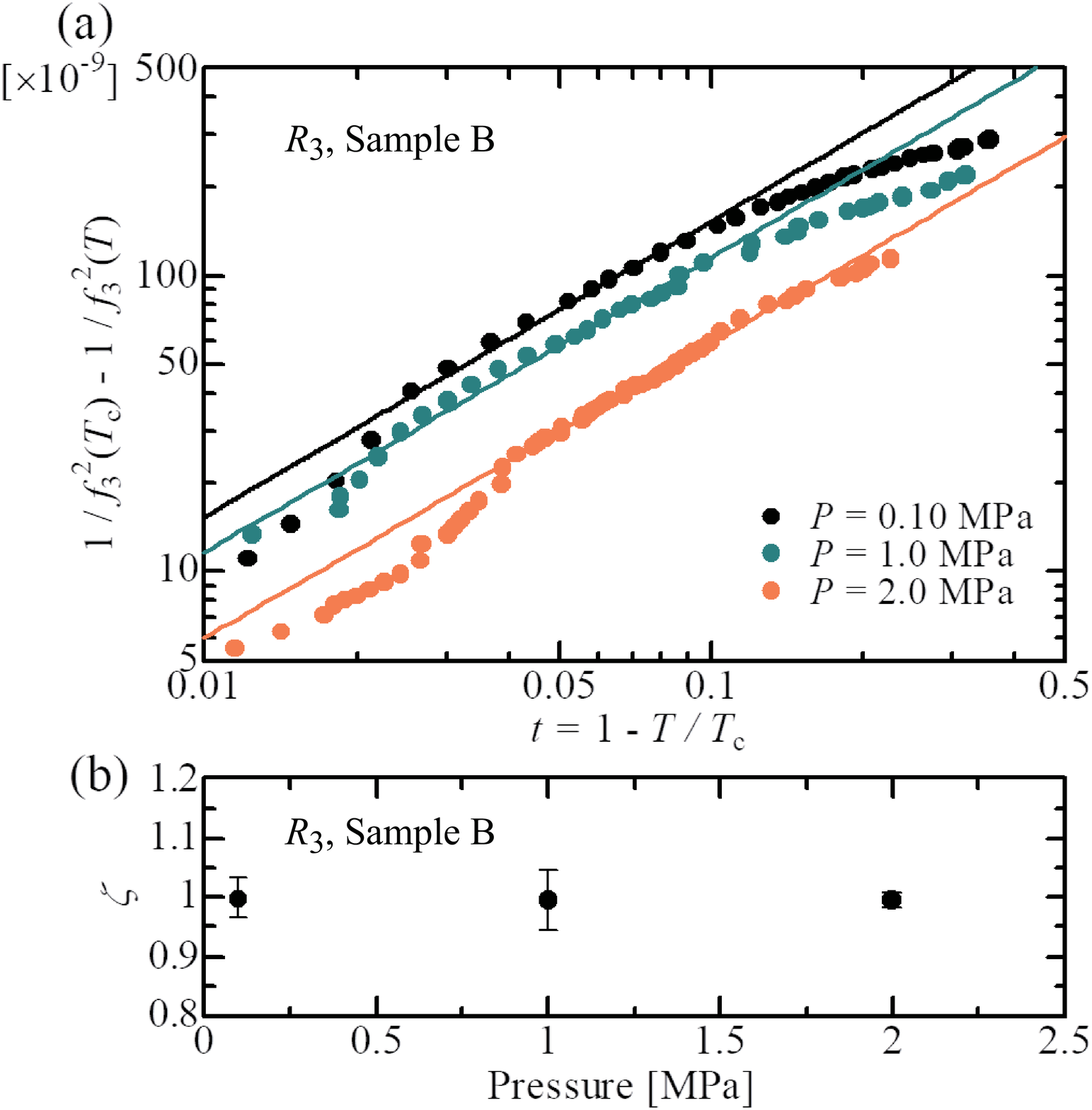}
\caption{(Color online)\label{R3CriticalExponent} Log-log plots of $1/f_{3}^{2}(T_{\rm c}) - 1/f_{3}^{2}(T)$ as a function of $t = 1 - T/T_{\rm c}$ for the $f_3$ data shown in Fig. \ref{R3FreqQinv} (a). 

}
\end{figure}

We analyze the powerlaw behavior of $\rho_{\mathrm s}$ near $T_{\mathrm c}$ by plotting $f_3^{-2}(T_{\mathrm c}) - f_3^{-2}(T)$ as a function of $t = 1 - T/T_{\mathrm c}$. 
The results are shown in Fig. \ref{R3CriticalExponent} (a).
Although the data are scattered, $f_3^{-2}(T_{\mathrm c}) - f_3^{-2}(T)$ can also be fitted by a power law in the temperature range $0.02 < t < 0.1$. 
As shown in Fig. \ref{R3CriticalExponent} (b), the exponent $\zeta$ is $1.00 \pm 0.05$ for three pressures.
It is concluded that the critical exponents obtained from the $R_3$ mode are consistent with those from the $R_2$ mode. 
It should, however, be emphasized that the $R_3$ mode is not fully identified, so that the formula, Eq. (\ref{f2}), deriving $\rho_{\mathrm s}$ has not been validated for the $R_3$ mode. 

We finally note that the dissipation shown in Fig. \ref{R3FreqQinv} (b) increases below $T_{\mathrm c}$ with a broad peak. 
It can be understood by the same phase matching mechanism as the dissipation in the $R_2$ mode.


\section{The correlation length in 4D}\label{Josephsonrelation}
The definition of the correlation length $\xi_{(0)}$ given by Eq. \ref{xi03D} is for the 3D superfluid. 
The definition of $\xi_{(0)}$ for arbitrary dimensions is given by\cite{FisherBarberJasnowPRA1973,Zassenhaus} 
\begin{equation}\label{xi04D}
\xi_{\mathrm (0)}(d) = \left( \frac{ k_{\mathrm B} T m^2_{\mathrm {^4He}} } { \hbar^2 \rho_{\mathrm {s(0)}} }\right)^{1/(d-2)}.
\end{equation}
where $d$ is the dimension of the system.
For $d=4$, the superfluid density $\rho_{\mathrm {s(0)}}$ must be defined in a 4D space including the imaginary time dimension, with a (SI) unit kg/m$^4$. 
In this Appendix, we show that $\xi_{(0)}(4)$ in 4D is nearly equal to $\xi_{(0)}(3)$, although there are some ambiguities in estimating the length scale in the imaginary time dimension.

Here we estimate $\xi_{(0)}(4)$ using the data taken at $P = 0.1$ MPa. 
In Sec. \ref{superfluidtransitioncorrelationlength}, we have obtained the zero-temperature limit of the superfluid density $\rho_{\mathrm {s(0)}}(3)$ to be $249$ kg/m$^3$. 
Here the number 3 in parentheses indicates 3D. 
In order to obtain the density in 4D space, one needs to know ``the lattice constant'' in the imaginary time dimension. 
The lattice constant $\delta\tau$ in unit of time is related to the ultraviolet cutoff energy $E_{\mathrm c}$ by $\delta\tau = \hbar/E_{\mathrm c}$\cite{SondhiRMP1997}. 
This is converted to the lattice constant $a_\tau$ in unit of length by multiplying the velocity of collective excitation $c$, i.e. $a_\tau = c\delta\tau$. 
In superfluid helium, $c$ is the phonon, i.e. sound, velocity, and the corresponding cutoff energy $E_{\mathrm c}$ should be the maximum phonon frequency $\hbar\omega_{\mathrm {ph}}$. 
We assume that $\omega_{\mathrm {ph}}$ is the phonon energy at the end of the linear dispersion in the well-known phonon - (maxon -) roton dispersion ($\omega - k$) curve. 
In the phonon region ($k < 1.0$ \AA$^{-1}$), the dispersion curve of $^4$He in Gelsil is identical to that of bulk superfluid $^4$He\cite{PlantevinPRB2002}. 
We therefore adopt $\omega_{\mathrm {ph}} = 1$ meV and $c = 240$ m/s at $P = 0.1$ MPa. 
Using these values, $a_\tau$ is estimated to be 0.16 nm. 
Then the superfluid density in 4D is obtained to be 
\begin{equation}
\rho_{\mathrm {s(0)}}(4) = \frac{\rho_{\mathrm {s(0)}}(3)}{a_\tau} =1.6 \times 10^{12} ~{\mathrm {kg/m}}^4.
\end{equation} 
The 4D correlation length $\xi_{\mathrm (0)}(4)$ is finally estimated to be 0.23 nm. 
This value is about the same as the 3D one, $\xi_{(0)}(3) \sim 0.33$ nm.
Considering the ambiguities in the physical quantities in these estimations, we conclude that the correlation length scale is identical in 3D and 4D. 

\section{Coupling of second sound standing waves with the mechanical resonator}\label{Secondsound}
As shown in Figs. \ref{freq_f_2} and \ref{dissipation_f_2}, we have observed anomalous behaviors in the dissipation $Q^{-1}$ and the corresponding frequency $f_2$, at temperatures between $T_{\lambda}$ and $T_{\mathrm c}$. 
For example, in the middle panel of Fig. \ref{dissipation_f_2} ($P = 1.0$ MPa), $Q^{-1}$ shows two peaks at $T \sim 1.9$ and 1.8 K, below which an excess amount of dissipation is observed down to 1.4 K, around which there is another peak. 
Correspondingly, $f_2$ shown in Fig. \ref{freq_f_2} shows small kinks, which coincide with the $Q^{-1}$ peaks at 1.8 K and at 1.4 K. 
Moreover, $Q^{-1}$ at different pressures have similar peak structures, which systematically shift to low temperatures with increasing pressure. 

We attribute these anomalies to the couplings of the $R_2$ resonance to standing wave modes of superfluid second sound.  
The excitation of second sound needs oscillation of temperature by local heating\cite{EnssHunklinger}. 
There are two possible sources:  
One is a heating by the loss of displacement current between the fixed electrode and flexible diaphragm oscillating at $f_{2}$. 
Another heat source is dissipation in the mechanical oscillation of the diaphragm. 
In the former, the heating is proportional to the square of the displacement current, $I_{\mathrm d}^{2}$, which oscillates at a frequency twice the mechanical resonant frequency, i.e. $2 f_{2}$, while in the latter, the heat oscillates with the same frequency $f_{2}$. 
Both of the heatings are estimated to be about 1 nW, which is enough to excite the second sound standing waves in the resonator. 
\begin{figure}[tb]
\centering
\includegraphics[width=1\linewidth]{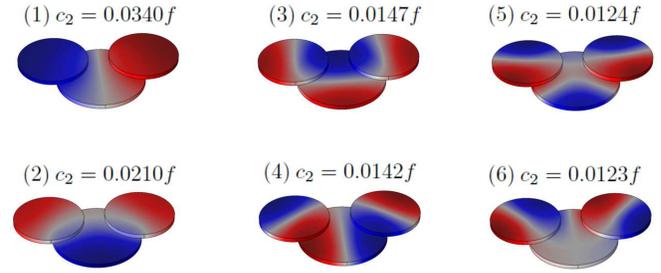}
\caption{(Color online) \label{coupling} 
Sound pressure distributions obtained from an FEM analysis of the standing waves with second sound velocity $c_{2}$ in the inner volume RI of mechanical resonator. 
The darkest red and the darkest blue correspond to the sound pressure 1.5 and -1.5 Pa, respectively.
Two disk-shaped volumes facing each diaphragm are connected to the main disk-shaped volume located at the center. 
Numbers in parentheses ((1) - (6)) indicate the first to sixth modes counted from the lowest frequency.
Each equality $c_2 = \lambda f$ determines the resonant frequency $f$. 
}
\end{figure}

\begin{figure}[tb]
\centering
\includegraphics[width=0.6\linewidth]{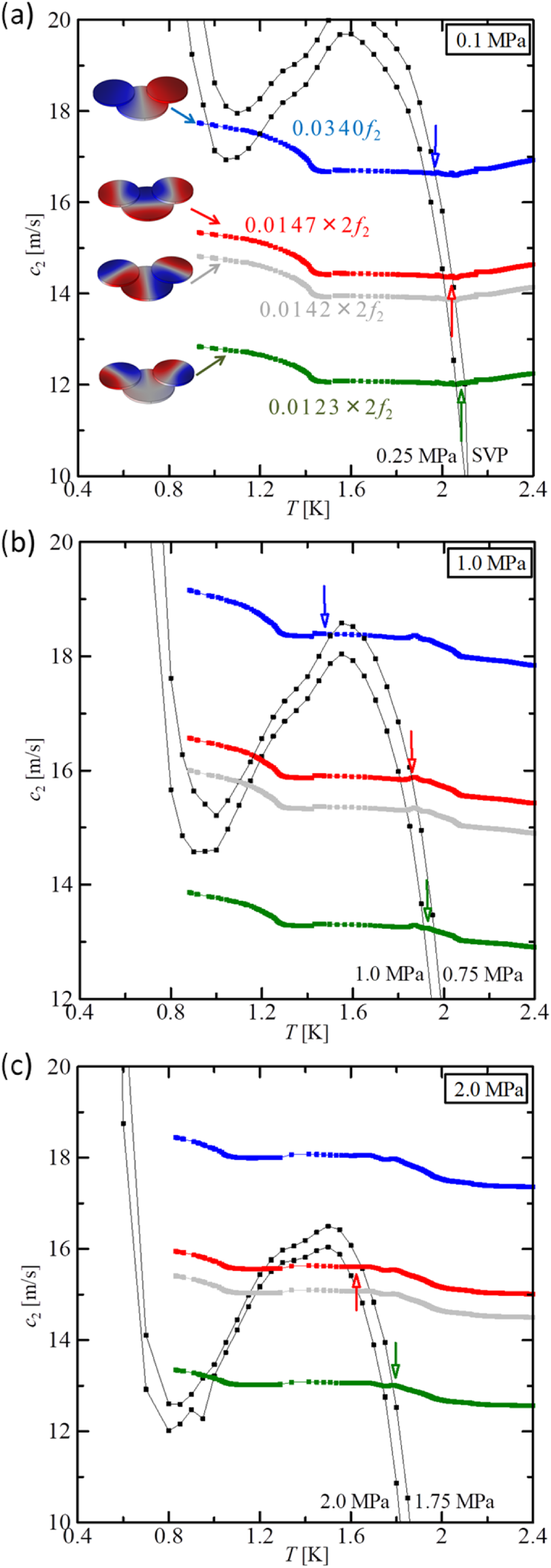}
\caption{(Color online) \label{2ndsoundcoupling} 
Crossing of the second sound standing waves and the mechanical resonance. 
Graphs show the data at (a) $P=0.1$, (b) 1.0, (c) 2.0 MPa. 
The black points show the second sound velocity $c_2$ taken from the data sheet of the second sound velocity at two different pressures that are close to the experimental pressures in this work\cite{BrooksDonnellyJPCRD1977}. 
Blace solid lines are the linear interpolation lines for guide to the eye.  
The $R_2$ resonance frequency $f_2(T)$ is converted to velocity using the relation $c_{\mathrm {res}} = \lambda f_2$ with a coefficient $\lambda$ determined by the FEM analysis. 
Colors of the data indicate different $\lambda$: 
Blue: 0.0340, red: 0.0147, gray: 0.0142, green: 0.0123.
When $c_2$ crosses to $c_{\mathrm {res}}$, i.e. the colored data points (indicated by colored arrows), the second sound can be excited by the mechanical resonator, resulting in an anomaly in $f_2$ accompanied with a peak in dissipation.
}
\end{figure}

It is difficult to calculate analytically calculate the fundamental modes of second sound resonances due to structural complexity of the mechanical resonator. 
Thus, we apply the finite element method (FEM) analysis to the resonator using COMSOL Multiphysics. 
Here we have performed a simple acoustic analysis of an ordinary density wave with the known second sound velocity to obtain the relationship between the sound velocity and the frequencies of the standing waves. 
Although the second sound is a temperature wave with no net pressure (or density) oscillation, the present method will also work to find the second sound modes which are determined only by the shape of the resonator and the sound velocity.
The results are shown in Fig.~\ref{coupling} as distributions of sound pressure. 
Each mode is represented just at the condition of a standing wave resonance. 
The condition is expressed as $c_{2} = \lambda f$, where $c_{2}$ is the (second) sound velocity, $f$ the frequency of standing wave and $\lambda$ the coefficient depending on the mode (having the dimension of wavelength). 
Note that the heating induced by displacement current produces oscillation at $f = 2 f_{2}$, while the heating by mechanical oscillation results in $f = f_{2}$. 

We have compared the frequencies of various second sound modes with $f_{2}(T)$ and have succeeded to assign the dissipation anomalies to the modes. 
The assignment of the modes is seen in Fig. \ref{2ndsoundcoupling}. 
In this figure, we plot $c_2$ with the ``mode velocity'', which is derived by using the formula $c_{\mathrm {res}} = \lambda f_2$ with coefficients $\lambda$ determined by the FEM analysis. 
When $c_{\mathrm {res}}$ crosses to $c_2$, the second sound can be excited by the mechanical resonator.
The coupling may produce an anomaly in $f_2$ accompanied with a peak in dissipation. 
Although $c_2$ data do not exactly correspond to the experimental pressure, we find some anomalies in $f_2$ near the crossing points, some of which are indicated by arrows. 

We identify that the first mode is excited at $f_{2}$, while the others are at $2 f_{2}$. 
The fourth mode seems not to be excited. 
The fifth and sixth modes can not be distinguished due to almost the same excitation conditions, although they can be simultaneously excited. 
In the data of $P=0.1$ MPa in Fig. \ref{2ndsoundcoupling} (a), the dissipation anomalies are assigned to be the first, third and sixth modes from the low temperature side. 
At $P = 1.0$ MPa (Fig. \ref{2ndsoundcoupling} (b)), the modes are the same as in the case of 0.1 MPa, but the first mode couples at the temperature where $c_2$ shows maximum and has little temperature dependence.
This results in broad dissipation peak around 1.4 K, as shown in Fig.~\ref{dissipation_f_2}. 
At $P = 2.0$ MPa (Fig. \ref{2ndsoundcoupling} (c)), the third and sixth modes are coupled with the resonance, and the third mode couples at the temperature of the maximum of $c_{2}$. 
This also produces a broad peak around 1.6 K. 

Near 1 K, i.e. below $T_{\mathrm c}$, $c_{2}$ takes a minimum depending on pressure\cite{MaurerPhysRev1951}. 
The third mode at 1.0 MPa and the sixth mode at 2.0 MPa may couple with the $R_2$ resonance around this minimum near $T_{\rm c}$. 
These couplings probably produce an additional dissipation that should have a very broad peak around 1 K. 
It is strongly suggested that the deviation between experimental data and the calculated dissipation observed at $P = 1.0$ and 2.0 MPa (Fig. \ref{dissipationanomaly} (b)) is caused by this coupling. 
At $P = 0.1$ MPa, no discrepancy between the dissipation data and the calculation. 
The absence of discrepancy at low pressures is also explained by the absence of the standing wave modes in the corresponding frequency regime. 
We therefore conclude that the phase matching process plus the second sound coupling account for the observed dissipation below $T_{\mathrm c}$ at all pressures. 

\section{Superfluid velocity}\label{Amplitude}
\begin{figure}[tb]
\centering
\includegraphics[width=0.9\linewidth]{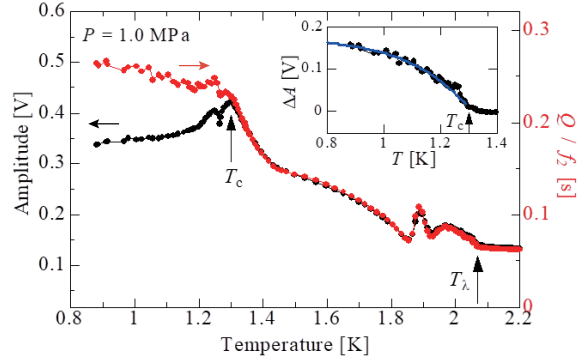}
\caption{(Color online) \label{amplitude} Temperature dependences of amplitude (black) for sample B under 1.0 MPa compared to the quality factor (red). Inset shows the deviation of the amplitude from the behavior of dissipation as a function of temperature. Blue solid line represents the $T$ dependence of superfluid density $\rho_{\rm s}$. Superfluid transition temperatures in bulk and in Gelsil are indicated by $T_{\rm \lambda}$ and $T_{\rm c}$}
\end{figure} 
 In this Appendix, we derive the superfluid velocity from the oscillation amplitude voltage $A$ of the resonator. 
% temperature dependence of the oscillation amplitude $A$ also exhibits a characteristic behavior. 
In Fig.~\ref{amplitude}, we show the temperature dependence of the amplitude $A(T)$ for sample B under 1.0 MPa, together with the quality factor devided by the resonant frequency, $Q(T)/f_2(T)$. 
In a usual condition, the amplitude $A$ corresponds to $Q$ by a relation of $Af_{\rm 0} \propto Q$, where $f_{0}$ is a resonant frequency. 
As is clearly seen in the graph, the amplitude $A$ matches exactly with $Q/f_{2}$ above $T_{\rm c}$. 
Below $T_{\rm c}$, however,  $A(T)$ deviates from $Q(T)/f_{2}(T)$. 
This deviation is attributed to the occurrence of superflow through Gelsil nanopores. 
Below $T_{\rm c}$, the amplitude of the pressure swing by the AC drive is relaxed due to the superflow between RI and RO, resulting in the decrease in the displacement of the detector diaphragm.

The displacement of the detector diaphragm D2, $x_{\mathrm {D2}}$, is related to $A$ by
\begin{equation}
  x_{\mathrm {D2}} = \frac{\epsilon A_{\mathrm {D2}}}{V_{\rm b}C^{2}}\frac{A}{2\pi Gf_{2}},
\end{equation}
where $\epsilon$ is the electric permittivity, $A_{\rm D2}$ is the area of the detector diaphragm, $V_{\rm b}$ = 350 V is the bias DC voltage applied to D2, $C$ is the capacitance in equilibrium position between D2 and the opposed electrode E2, and $G$ = 500 MV/A is the gain of the current preamplifier. 
The amplitude $A$ directly corresponds to the mass current of liquid $^4$He moving in RI and flowing through Gelsil. 
The reduction in $A$ attributed to mass current through Gelsil, $\Delta A$, is estimated as the deviation of $A$ from the temperature dependence of $Q/f_{2}$ shown in Fig.~\ref{amplitude}. 
Then $\Delta A$ determines mass current $J_{\rm s}$ by the following relation
\begin{equation}
  \rho A_{\rm {D2}}\frac{dx}{dt} = J_{\rm s} = \rho_{\rm s} v_{\rm s} S,
\label{vs}
\end{equation}
where $v_{\rm s}$ is the superfluid velocity in Gelsil and $S$ the total cross sectional area of the flow channel in Gelsil. 
Unless the superfluid velocity $v_{\rm s}$ exceeds a critical velocity, $v_{\rm s}$ is determined by the chemical potential difference across the channel. 
The inset of Fig.~\ref{amplitude} shows the temperature dependence of $\Delta A$.
It is remarkable that $\Delta A$ matches with the temperature dependence of superfluid density $\rho_{\rm s}(T)$ estimated from the $R_1$ mode frequency $f_{1}$, as indicated by the blue line. 
It is therefore concluded that $v_{\rm s}$ is constant and $\Delta A$ is well explained by Eq.~(\ref{vs}). 
This analysis provides a strong support to our hydrodynamic model of the resonator and the validity of the determination of $\rho_{\rm s}$ by Eq.~(\ref{f2}) and Eq.~(\ref{f1}). 
The superfluid velocity estimated from Eq.~(\ref{vs}) is $v_{\rm s} \simeq 2$ mm/s.


\begin{thebibliography}{99}
\bibitem{SondhiRMP1997} S. L. Sondhi, S. M. Girvin, J. P. Carini, and D. Shahar, Rev. Mod. Phys. {\bf 69}, 315 (1997). 
\bibitem{SachdevQPT} S. Sachdev, Quantum Phase Transitions, Cambridge University Press (2011).
\bibitem{GegenwartNP2008} P. Gegenwart, Q. Si, and F. Steglich, Nature Phys. {\bf 4}, 186 (2008).
\bibitem{KeimerN2015} B. Keimer, S. A. Kivelson, M. R. Norman, S. Uchida, and J. Zaanen, Nature {\bf 518}, 179 (2015).
\bibitem{BrounPRL2007} D. M. Broun, W. A. Huttema, P. J. Turner, S. \"{O}zcan, B. Morgan, R. Liang, W. N. Hardy, and D. A. Bonn, Phys. Rev. Lett. {\bf 99}, 237003 (2007)
\bibitem{FranzPRL2006} M. Franz and A. P. Iyengar, Phys. Rev. Lett. {\bf 96}, 047007 (2006).
\bibitem{YamamotoPRL2004} K. Yamamoto, H. Nakashima, Y. Shibayama and K. Shirahama, Phys. Rev. Lett. {\bf 93} 075302 (2004).
\bibitem{YamamotoPRL2008} K. Yamamoto, Y. Shiabayama and K. Shirahama, Phys. Rev. Lett. {\bf 100} 195301 (2008).
\bibitem{ShirahamaJLTP2007} K. Shirahama, J. Low Temp. Phys. {\bf 146}, 485-497 (2007).
\bibitem{ShirahamaLTP2008} K. Shirahama, K. Yamamoto and Y. Shibayama, Low Temp. Phys. {\bf 34}, 273 (2008).
\bibitem{ShirahamaJPSJ2008} K. Shirahama, K. Yamamoto and Y. Shibayama, J. Phys. Soc. Jpn. {\bf 77}, 111011 (2008).
\bibitem{EggelPRB2011} Th. Eggel, M. Oshikawa and K. Shirahama. Phys. Rev. B {\bf 84}, 020515(R) (2011).
\bibitem{EggelThesis2011} Th. Eggel, Ph.D. Thesis, University of Tokyo (2011).
\bibitem{AhlersRMP1980} D. S. Greywall and G. Ahlers, Phys. Rev. A {\bf 7}, 2145 (1973); 
G. Ahlers, Rev. Mod. Phys. {\bf 52}, 489 (1980).
\bibitem{BarmatzRMP2007} M. Barmatz, I. Hahn, J. A. Lipa, and R. V. Duncan, Rev. Mod. Phys. {\bf 79}, 1 (2007).
\bibitem{FisherPRB1989} M. P. A. Fisher, P. B. Weichman, G. Grinstein, and D. S. Fisher, Phys. Rev. B {\bf 40}, 546 (1989).
\bibitem{AvenelPRL1985} O. Avenel and E. Varoqaux, Phys Rev. Lett. {\bf 55}, 2704 (1985).
\bibitem{RojasPRB2015} X. Rojas and J. P. Davis, Phys. Rev. B {\bf 91}, 024503 (2015).
\bibitem{TaniJPSJ2021} T. Tani, Y. Nago, S. Murakawa, and K. Shirahama, J. Phys. Soc. Jpn. {\bf 90}, 033601 (2021).
%\bibitem{PlantevinPRB2002} O. Plantevin, H. R. Glyde, B. Fak, J. Bossy, F. Albergamo, N. Mulders, and H. Schober, Phys. Rev. B {\bf 65} 224505 (2002).
\bibitem{BarrettJACS1951} E. P. Barrett, L. G. Joyner, and P. P. Halenda, J. Am. Chem. Soc. {\bf 73}, 373 (1951).
\bibitem{BlaauwgeersJLTP2007} R. Blaauwgeers, M. Blazkova, M. Clivecko, V. B. Eltsov, R. de Graaf, J. Hosio, M. Krusius, D. Schmoranzer, W. Schoepe, L. Skrbek, P. Skyba, R. E. Solntsev, and D. E. Zmeev, J. Low Temp. Phys. {\rm 146}, 537 (2007).
\bibitem{GoldnerJLTP1993} L. S. Goldner, N. Mulders, and G. Ahlers, J. Low Temp. Phys. {\bf 93}, 131 (1993).
%\bibitem{LeGuillouPRL1977} J. C. Le Guillou and J. Zinn-Justin, Phys. Rev. Lett. {\bf 39}, 95 (1977); Phys. Rev. B {\bf 21}, 3976 (1980).
\bibitem{CampostriniPRB2001} M. Campostrini, M. Hasenbusch, A. Pelissetto, P. Rossi, and E. Vicari, Phys. Rev. B {\bf 63}, 214503 (2001).
\bibitem{LipaPRB2003} J. A. Lipa, J. A. Nissen, D. A. Stricker, D. R. Swanson, and T. C. P. Chui, Phys. Rev. B{\bf 68}, 174518 (2003).
\bibitem{KiewietPRL1975} C. W. Kiewiet, H. E. Hall and J. D. Reppy, Phys. Rev. Lett. {\bf 35}, 1286 (1975).
\bibitem{YoonPRL1997} J. Yoon and M. H. W. Chan, Phys. Rev. Lett. {\bf 78}, 4801 (1997).
\bibitem{HarrisJPC1974} A. B. Harris, J. Phys. C {\bf 7}, 1671 (1974).
\bibitem{ChanMuldersReppyPhysToday1996} M. Chan, N. Mulders, and J. Reppy, Phys. Today {\bf 49} (8), 30 (1996).
\bibitem{ChanPRL1988} M. H. W. Chan, K. I. Blum, S. Q. Murphy, G. K. S. Wong and J. D. Reppy, Phys. Rev. Lett. {\bf 61} 1950 (1988). 
\bibitem{WongPRB1993} G. K. S. Wong, P. A. Crowell, H. A. Cho, and J. D. Reppy, Phys. Rev. B {\bf 48}, 3858 (1993); G. K. S. Wong, Ph.D. Thesis, Cornell University (1990). 
\bibitem{YoonPRL1998} J. Yoon, D. Sergatskov, J. Ma, N. Mulders, and M. H. W. Chan, Phys. Rev. Lett. {\bf 80}, 1461 (1998).
\bibitem{MuldersPRL1991} N. Mulders, R. Mehrotra, L. S. Goldner, and G. Ahlers, Phys. Rev. Lett. {\bf 67}, 695 (1991).
\bibitem{NishimoriOrtiz} H. Nishimori and G. Ortiz, Elements of Phase Transitions and Critical Phenomena, OUP Oxford (2010).
\bibitem{DonnellyBarenghi1998} R. J. Donnelly and C. F. Barenghi, J. Phys. Chem. Ref. Data {\bf 27}, 1217 (1998).
\bibitem{Zassenhaus} G. Zassenhaus, Ph. D. Thesis, Cornell University, 1999.
\bibitem{BrewerJLTP1970} D. F. Brewer, J. Low Temp. Phys. {\bf 3}, 205 (1970).
\bibitem{SyskakisPRL1985} E. G. Syskakis, F. Pobell, and H. Ullmaier, Phys. Rev. Lett. {\bf 55}, 2964 (1985).
\bibitem{GaspariniRMP2008} F. M. Gasparini, M. O. Kimball, K. P. Mooney, and M. Diaz-Avila, Rev. Mod. Phys. {\bf 80}, 1195 (2008).
\bibitem{JosephsonPhysLett1966} B. D. Josephson, Phys. Lett. {\bf 21}, 608 (1966).
\bibitem{FisherBarberJasnowPRA1973} M. E. Fisher, M. N. Barber, and D. Jasnow, Phys. Rev. A {\bf 8}, 1111 (1973).
\bibitem{ZassenhausPRL1999} G. Zassenhaus and J. D. Reppy, Phys. Rev. Lett. {\bf 83}, 4800 (1999).
\bibitem{MakiuchiPRB2018} T. Makiuchi, M. Tagai, Y. Nago, D. Takahashi, and K. Shirahama, Phys. Rev. B {\bf 98}, 235104 (2018).
\bibitem{GorJCP2015} G. Y. Gor, D. W. Siderius, C. J. Raumussen, W. P. Krekelberg, V. K. Shen and N. Bernstein, J. Chem. Phys. {\bf 143}, 194506 (2015).
\bibitem{WeinribHalperin} A. Weinrib and B. I. Halperin, Phys. Rev. B {\bf 27}, 413 (1983).
%\bibitem{YamamotoD} K. Yamamoto, Ph. D. Thesis, Keio University (2008).
\bibitem{JosephsonPhysLett1962} B. D. Josephson, Phys. Lett. {\bf 1}, 251 (1962).
\bibitem{AndersonRMP1966} P. W. Anderson, Rev. Mod. Phys. {\bf 38}, 298 (1966).
\bibitem{PackardRMP1998} R. E. Packard, Rev. Mod. Phys. {\bf 70}, 641 (1998).
\bibitem{KobayashiJPSJ2010} T. Kobayashi, J. Taniguchi, A. Saito, S. Fukazawa, M. Suzuki, K. Shirahama, J. Phys. Soc. Jpn. {\bf 79}, 084601 (2010).
\bibitem{YamamotoJPSJ2007} K. Yamamoto, Y. Shibayama and K. Shirahama. J. Phys. Soc. Jpn. {\bf 77}, 013601 (2007).
\bibitem{ProsniakSR2019} O. A. Pro\'{s}niak, M. \L\c{a}cki, and B. Damski, Sci. Rep. {\bf 9}, 8687 (2019).
\bibitem{MaurerPhysRev1951} R. D. Maurer and M. A. Herlin, Phys. Rev. {\bf 81}, 444 (1951).
\bibitem{PlantevinPRB2002} O. Plantevin, H. R. Glyde, B. F\aa k, J. Bossy, F. Albergamo, N. Mulders, and H. Schober, Phys. Rev. B {\bf 65}, 224505 (2002).
\bibitem{EnssHunklinger} See for example, \textit{Low Temperature Physics}, Ch. Enss and S. Hunklinger, Springer (2005).
\bibitem{BrooksDonnellyJPCRD1977} J. S. Brooks and R. J. Donnelly, J. Phys. Chem. Ref. Data {\bf 6}, 51 (1977).
%\bibitem{ChowPRL1998} E. Chow, P. Delsing, and D. B. Haviland, Phys. Rev. Lett. {\bf 81}, 204 (1998); D. B. Haviland, K. Anderson, P. {\AA}gren, J. Johansson, V. Sch\"{o}llman, and M. Watanabe, Physica C {\bf 352}, 55 (2001).
\end{thebibliography}
\end{document}